\documentclass[]{aastex701}

\begin{document}

\title{Radio and Optical Flares on the dMe Flare Star EV Lac}

\author[orcid=0000-0001-5643-8421]{Rachel A. Osten}
\altaffiliation{Johns Hopkins University Center for Astrophysical Sciences}
\affiliation{Space Telescope Science Institute}
\email[show]{osten@stsci.edu}

\author[0000-0001-7458-1176]{Adam F. Kowalski}
\affiliation{Department of Astrophysical and Planetary Sciences, University of Colorado Boulder, CO 80305, USA}
\altaffiliation{Laboratory for Atmospheric and Space Physics, Boulder, CO 80303, USA}
\altaffiliation{National Solar Observatory, Boulder, CO 80303, USA}
\email{Adam.F.Kowalski@Colorado.EDU}

\author[0000-0002-6629-4182]{Suzanne Hawley}
\affiliation{University of Washington, Seattle}
\email{slh@uw.edu}

\author[0000-0001-5974-4758]{Isaiah I. Tristan}
\affiliation{Rice Space Institute, Rice University, Houston, TX  77005, USA}
\affiliation{Department of Astrophysical and Planetary Sciences, University of Colorado Boulder, CO 80305, USA}
\altaffiliation{Laboratory for Atmospheric and Space Physics, Boulder, CO 80303, USA}
\altaffiliation{National Solar Observatory, Boulder, CO 80303, USA}
\email{Isaiah.Tristan@colorado.edu}

\author{Sarah J. Schmidt}
\affiliation{}
\email{sarahjaneschmidt@gmail.com}

\author[0000-0003-2053-0749]{Ben Tofflemire}
\affiliation{Department of Astronomy, The University of Texas at Austin, Austin, TX 78712, USA}
\email{tofflemire@utexas.edu}

\author{Eric Hilton}
\affiliation{}
\email{ericjhilton@gmail.com}
\setcounter{footnote}{1}

\begin{abstract}

We present the results of a coordinated campaign to observe radio and optical stellar flares from the nearby M dwarf flare star EV~Lac. 
From a total of 27 hours of radio  and 29 hours of optical observations, we examine 
the correspondence of the action of accelerated electrons of different energies in two distinct regions of the stellar atmosphere.
We find that out of 9 optical flares with suitable radio coverage, only four have plausible evidence for a radio response. 
Optical photometric properties cannot predict which flares will have a radio response.
From flares with time-resolved optical spectroscopy available, optical-only flares have similar implied electron distributions, while those with radio responses better correlate with higher low-energy cutoffs.
The optical flares with a radio response all exhibit a delay between the optical and radio peaks of $\approx$1-7 minutes, with the optical flare peaking earlier in all cases. This likely indicates multiple loops are involved in the event, and/or 
the different impacts on electrons trapped in a magnetic loop (producing radio emission), versus those directly precipitating from the loop (producing the optical flare).
We also remark on the radio spectral index behavior at early times for the largest radio flare observed in this study, which we interpret as evidence for increased opacity from a chromospheric evaporation front. 

\end{abstract}

\keywords{
\uat{Red dwarf flare stars}{1367} ---
\uat{Discrete radio sources}{389} --- 
\uat{Stellar flares}{1603} ---
\uat{Variable stars}{1761}
}


\section{Introduction} 
Stellar flares are stochastic brightening events that occur on stars in the cool half of the  
Hertzsprung-Russell diagram 
\citep{YangLiu2019}
as a result of magnetic reconnection and energy release \citep{benzgudel2010araa}.
Briefly, energy is released as a result of magnetic field motion and subsequent reconnection.  This energy goes into particle acceleration, plasma heating, and mass motions \citep{Kowalski2024}. It involves all atmospheric layers of the star, and because of the different physical processes involved, produces emissions across the electromagnetic spectrum. 

There are a few key criteria which predict the types of stars on which frequent and/or extreme flares will occur, and these can be summarized into two broad categories of rotation and internal structure. 
Rapidly rotating stars 
are known to display enhanced magnetic activity, which includes flaring, vis. flares on young stars \citep{McClearyWolk2011}, tidally locked binaries \citep{ostenetal2004}, and flaring on stars experiencing rotation enhancements \citep{Ayres2001,testa2007}.  
Flares also occur on stars with outer convection zones, most notably in M dwarfs where energy transfer in the interior takes place either mostly or fully  via convection. 
\citet{Osten2016} summarizes flare activity on different types of stars and at different ages. 

M dwarfs are the most common type of star locally \citep{HenryJao2024}, and they fulfill at least one of the criteria listed above. 
Young M dwarfs are expected to have enhanced flare rates, and this is observed \citep{Tristan2023}. 
They are also projected to be the most common type of planet-hosting star
\citep{Yang2020}.
Recent studies have suggested both positive \citep{Rimmer2018,Chen2021} and negative \citep{Feinstein2022} outcomes of flares for any close-in planets. 
While this aspect of flare studies plays out, M dwarf flares remain excellent candidates for placing the well-studied solar flares in perspective with the majority of nearby flaring stars and learning about the relative contribution of physical processes such as particle acceleration in different types of stars.

Optical observations show the response of the lower atmosphere of the star to the energy input occurring during a stellar flare and have a long history in the study of flaring on M dwarfs \citep{Hertzsprung1924,vanMaanen1940}. 
As a result of improvements to high cadence, low spectral resolution optical time-resolved spectroscopy, \citet{Kowalski2013} (hereafter K13) revealed insights into the changing blue-optical spectral energy distribution (SED) during the rise, peak, and decay phases of M dwarf flares.
The  description of this SED has evolved from one with the shape of a blackbody \citep{Mochnacki1980}
to radiative hydrodynamical models detailing the response of the atmosphere to an input of a spectrum of energetic particles which precipitate directly from a magnetic trap before impinging the lower atmosphere
\citep{KAC2024}. 
Recent results \citep{Kowalski2025} have shown that the 
near-ultraviolet (NUV) spectrum of two intense flares was indicative of a rising spectrum into the ultraviolet (UV), requiring the existence of electrons with energies above 500 keV to explain the SED. 
This demonstrates that particle acceleration plays an important role in controlling the flare emission in the UV through optical wavelength regions, and that spectroscopy can reveal crucial information about the 
electron distribution.

Radio observations of stellar flares provide direct evidence of the existence and characteristics of accelerated particles in stellar atmospheres. 
Incoherent gyrosynchrotron radiation from a power-law distribution of electrons produces radio emission
at harmonics $s$ of the electron gyrofrequency \citep[$\nu_{B}=2.8\times10^{6} B (G) \; \rm{MHz}$;][]{dulk1985}, where $B$ is the magnetic field strength in Gauss in the radio-emitting source and $10<s<100$.
The quiescent radio emission from active stars at GHz frequencies as well as variable largely unpolarized flaring emission is generally attributed to 
gyrosynchrotron emission \citep{benzgudel2010araa} --
the nearly 100\% circularly polarized bursts, on the other hand, are attributed to a coherent mechanism like electron-cyclotron maser emission or plasma emission \citep{dulk1985}.
Source inhomogeneities and optical depth effects render the 
interpretation of flaring cm wavelength emission complex \citep{ostenetal2005}, although recent observations at mm wavelengths are revealing optically thin flare emission which enables constraints on the electron distribution \citep{MacGregor2020}, confirming results from solar flares for a distribution with relatively more high than low energy electrons.
Recent radio observing at frequencies higher than 10 GHz has generally been successful at observing some flares in the optically thin regime, although even in the 12-18 GHz range some flares are still optically thick \citep{Tristan2025}. 
Solar flare observations reveal that  microwave emission 
is produced both from electrons trapped in the magnetic loop as well as those which precipitate from the trap, a so-called ``trap plus precipitation" model
\citep{Kundu2001}.

Coordinated observations enable probes of how the physical processes involved in flares (e.g. particle acceleration, plasma heating) interrelate and add context and depth
to interpretation.
Since flares involve all layers of the atmosphere, there will be different responses in different parts of the electromagnetic spectrum. 
There has been a large emphasis in multi-wavelength observations on relating radio emission with coronal X-ray emission to explore the G\"{u}del-Benz relation \citep{gb1989} and Neupert effect \citep{gudel1996}, or 
optical and coronal emission
\citep[e.g.,][]{hawley1995,caballero2015}, but radio and optical/ultraviolet coordinated observations have been sparser.
\citet{gagne1998} reported on the results of a radio, optical, and X-ray multiwavelength campaign on the M dwarf binary EQ Peg AB, finding
numerous optical flares but most without an apparent increase at 3.5 cm (and those radio flares which did appear had high degrees of circular polarization, suggesting a coherent mechanism). 
\citet{smith2005} and \citet{mk2005} reported on X-ray and radio, and X-ray and NUV observations respectively, of the same sample of nearby M dwarf stars, with the X-ray observations from XMM-Newton in common.
Although the joint occurrence of radio and NUV flares was not discussed, inspection of the figures in the respective papers shows that for 8 X-ray flares with both NUV and radio data observed on 4 different M dwarfs, six flares exhibited an NUV response, of which only two showed an obvious microwave response. 
\citet{ostenetal2005} reported on a large flare from the flare star EV~Lac observed during radio, optical and X-ray measurements, finding that the optical flare peak occurred less than a minute before the radio peak. 
More recently, \citet{Tristan2025} explored mainly 12-18 GHz radio observations of the early M dwarf AU~Mic within the context of a large multi-wavelength campaign involving optical, UV and X-ray measurements. 
They found a  high rate of association between the radio flares and UV and/or optical flares.

Solar physicists generally use nonthermal bremsstrahlung emission originating at hard X-ray energies \citep{Krucker2008}   as a method for diagnosing the distribution of accelerated particles in flares, as inversion techniques can relate the observed X-ray photon spectrum to the particle spectrum \citep{brown1971} and enable a study of its variation with time during flares and between flares. 
Such techniques are out of reach for all but the most luminous stellar flares \citep{osten2007}\footnote{Even in these cases, the extraordinarily hot nature of the thermal plasma with temperatures $>$ 100 MK all but obscures any nonthermal response; \citep{Ostenetal2016}}, due to sensitivity limitations of astronomical hard X-ray detectors.
In the absence of that wavelength capability, the optical and radio regimes are the most promising for studies of flare-related particle acceleration in a stellar context.
The white light and radio wavelength ranges probe different particle energies, as well as different regions of the atmosphere.
It is largely particles with energies higher than a few hundred keV to MeV-level that contribute to gyrosynchrotron emission \citep{dulk1985}
The radio-emitting regions typically are 
in low density environments.
In contrast, white light flare measurements come from the heating from lower energy electrons, typically below about 100 keV \citep[][although note the extraordinary result cited above implicating 500 keV electrons]{KAC2024}, and much lower down in the atmosphere, at chromospheric levels.
Therefore a joint study of radio and optical flaring emission will produce complementary constraints on the particle population and the interaction of these particles with the ambient plasma at different layers of the atmosphere.

EV~Lac is a highly active mid-M dwarf known for its extreme flare outbursts across the electromagnetic spectrum, from radio \citep{ostenetal2005} to optical \citep{roizman1982} and hard X-ray wavelengths \citep{osten2010}. 
It is nearby \citep[d=5.05 pc,][]{GaiaDR3} with a spectral type of M3.5 
\citep{Reid1995}. 
The star's galactic velocity and position are consistent with an age of $\sim$400 MY, although the precise age is difficult to obtain \citep[see discussion in][]{Paudel2021}.
Its radius is approximately 0.35 R$_{\odot}$ \citep{Paudel2021}.
Among the handful of active M dwarfs whose photospheric magnetic configurations have been studied using Zeeman Doppler Imaging, EV~Lac stands out for having highly non-axisymmetric magnetic fields \citep{donati2009}, while most others with similar stellar parameters exhibit a high degree of axisymmetry.
\citet{Donati2025} recently reported on circularly and linearly polarized optical spectra of EV Lac along with least-squares deconvolution to probe the star's magnetic topology, finding it to be complex, strong overall and confirming the lower degree of axisymmetry than other active M dwarfs, with evidence for intense 6 kG small-scale fields when averaged over the stellar surface.
EV~Lac has been the subject of previous multi-wavelength flare campaigns, notably \citet{ostenetal2005,Paudel2021,Inoue2024}.

We present the results of a coordinated campaign to observe radio and optical stellar flares from the nearby M dwarf flare star EV~Lac taken over the course of 4 days. 
We examine a total of about 27 hours hours of radio observations and 29 hours of optical observations.
While much of the optical photometry and spectroscopy has been published in prior studies \citep{tofflemire2012,Kowalski2013}
the radio observations are presented here for the first time.
Even with higher sensitivity radio receivers and easier ability to obtain optical observations nowadays with TESS photometry, the present dataset is unique in its extensive coverage, high time resolution at optical wavelengths, and on one night, the presence of time-resolved optical spectroscopy. 
We also make use of newly available models to interpret the optical flare spectral energy distributions.
The paper is arranged as follows: \S~\ref{sec:observations}
describes the observations and radio data reductions,
\S~\ref{sec:analysis} describes the multi-frequency behavior of large radio flares, 
co-occurrence of radio and optical flares, and 
time-resolved optical spectroscopy.
\S~\ref{sec:disc} discusses the findings and
\S~\ref{sec:conc} concludes with implications for future multi-wavelength observations.

\section{Observations and Reductions} \label{sec:observations}


\begin{deluxetable}{llll}
\tablewidth{0pt}
\tablecolumns{4}
\tablecaption{Summary of Radio and Optical Observations \label{sumtbl}}
\tablehead{ \colhead{Date} & \colhead{Radio Obs. }& \colhead{Optical Obs.} & \colhead{notes}\\
\colhead{} & \colhead{Start$-$Stop (d)} & \colhead{Start$-$Stop (d)} & \colhead{} }
\startdata
2009 Sept. 20 &    20.10-20.42   & 20.16-20.50    & DAO:  spectroscopy at DAO            \\
 &            & \textemdash{}& APO: bad weather \\
2009 Sept. 21  &  21.10-21.42                  &      21.23-21.50                  &     APO: NMSU 1m U band photometry       \\
2009 Oct. 8     & 8.18-8.37 &     8.0-8.23  & KPNO:  WIYN 0.9m U band photometry\tablenotemark{1} \\
&                &     \textemdash{}                  &        APO: bad weather       \\
2009 Oct. 10   &   10.05-10.36                 &        10.06- 10.44             &    
APO: NMSU 1m U band photometry \tablenotemark{2}        \\
                &                               & 10.13-10.24 & APO: ARCSAT 0.5m u band photometry\tablenotemark{2} \\
& & 10.07-10.41& APO: SDSS low-resolution optical spectroscopy\tablenotemark{2}\\ 
\enddata
\tablenotetext{1}{published in \citet{tofflemire2012} }
\tablenotetext{2}{published in K13}
\end{deluxetable}

\subsection{Radio Observations\label{sec:radio_obs}}

Observations of EV~Lac were taken over the course of 4 days: those on 2009 Sept. 20 and 21 occurred when the VLA was in a reconfiguration between C and DnC arrays (moving to a more compact array), while the observations on 2009 Oct. 8 and 10 observations occurred in the hybrid DnC configuration.
The observations used 0137+331 as the flux calibrator, and 2255+420 as the phase calibrator. 
All of the VLA
observations were performed with two subarrays of roughly half the number of available antennas in each subarray. 
One subarray used the X-band receiver (3.6 cm) and the other subarray used the C-band receiver (6 cm).
The project code for the 2009 VLA observations is AO253.
Note the sensitivity of the
observations is pre-upgrade VLA, so only 100 MHz of bandwidth per polarization is available.  

\begin{figure*}[ht!]
\vspace*{-8cm}
\includegraphics[scale=0.3]{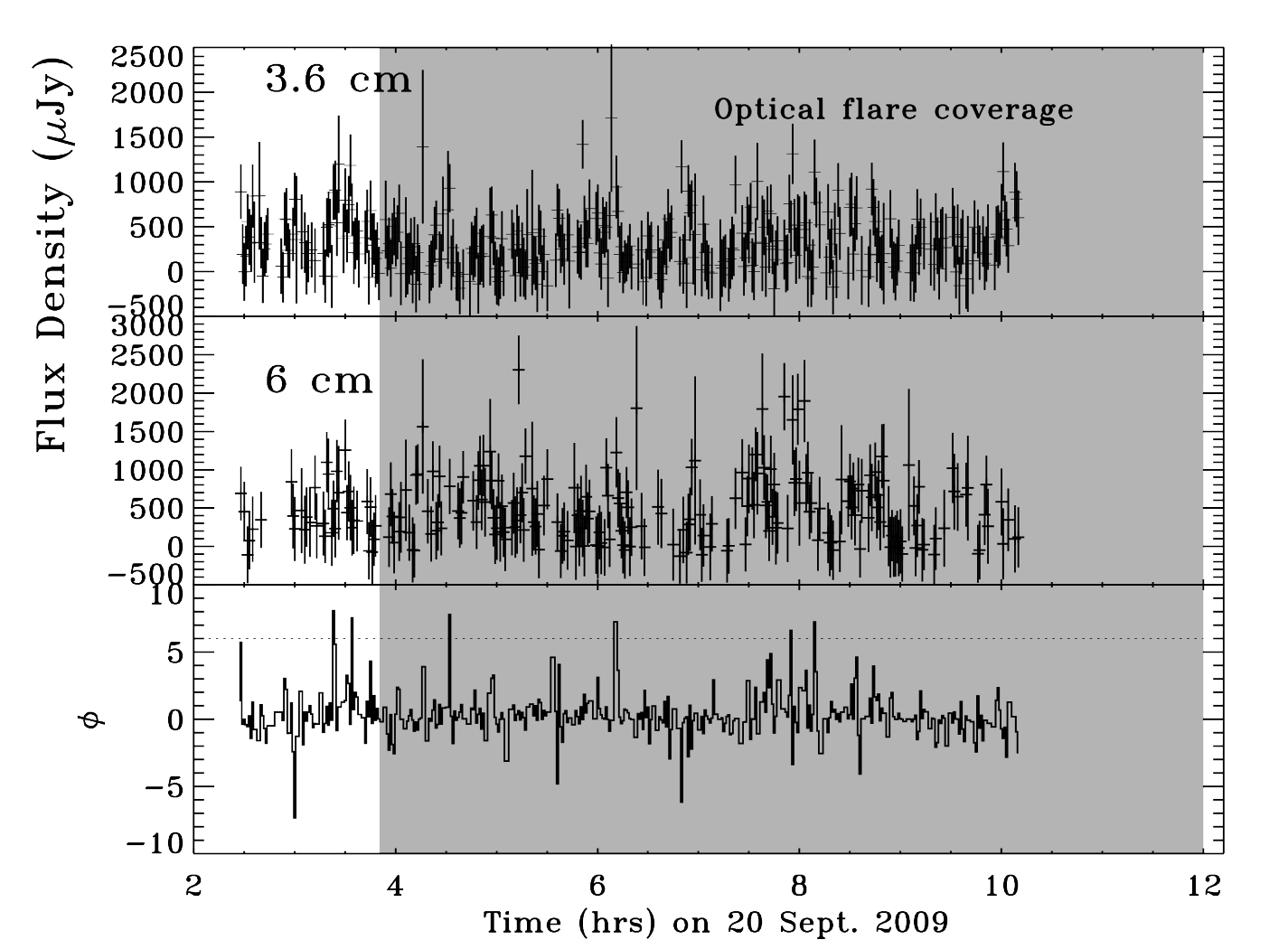}
\includegraphics[bb=0 0 1000 1800,clip,scale=0.3]{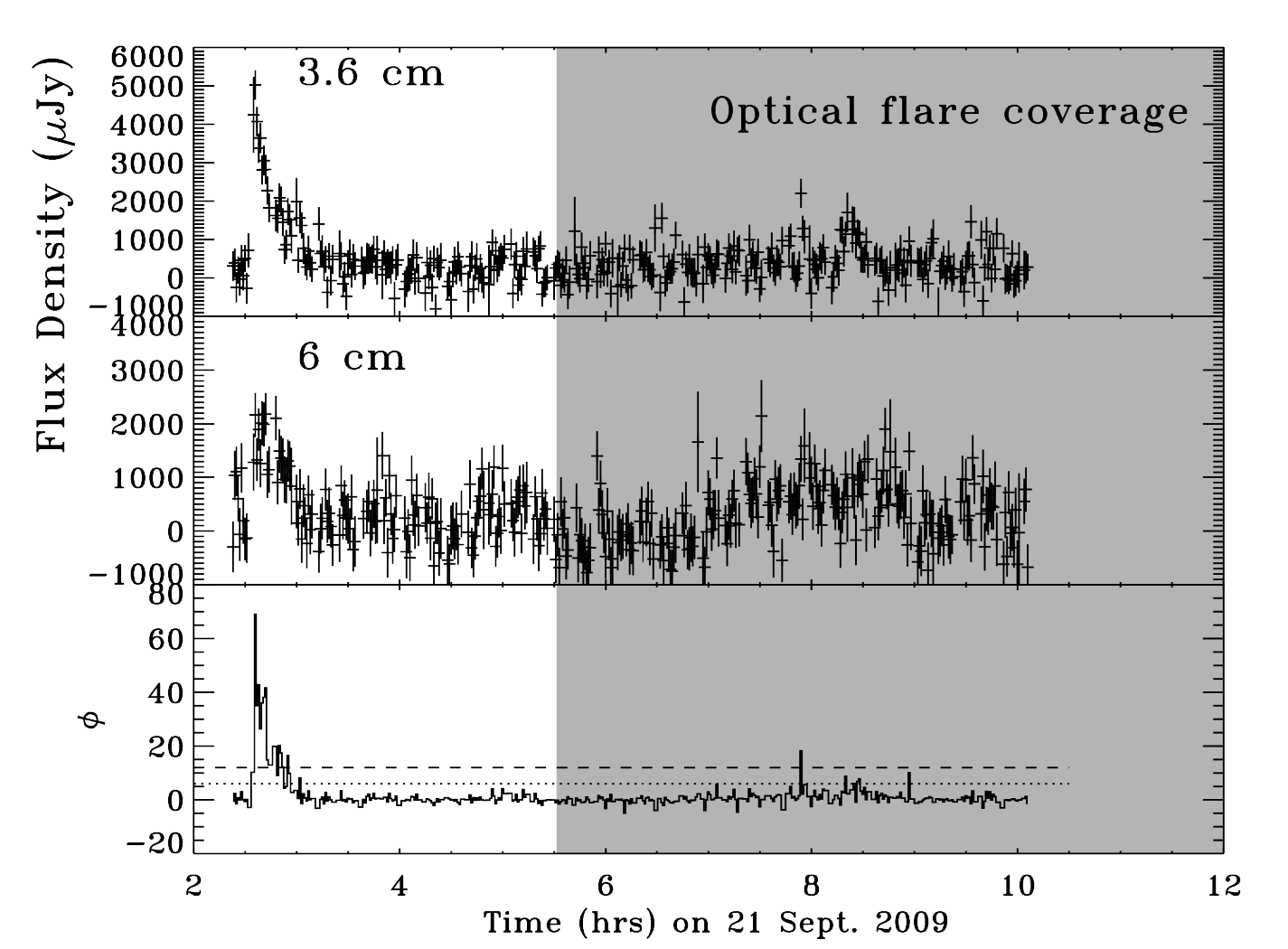}
\includegraphics[scale=0.3]{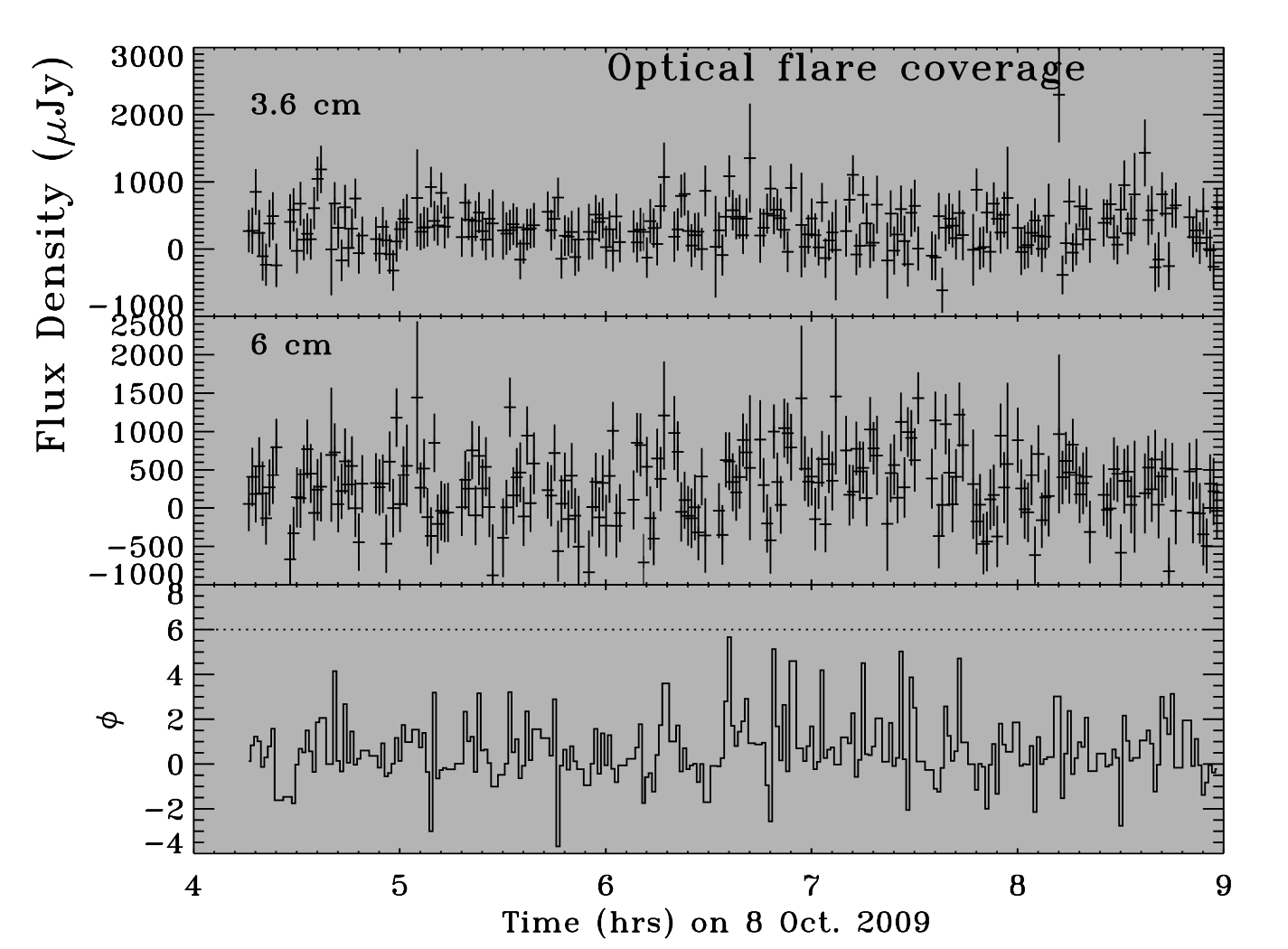}
\includegraphics[scale=0.3]{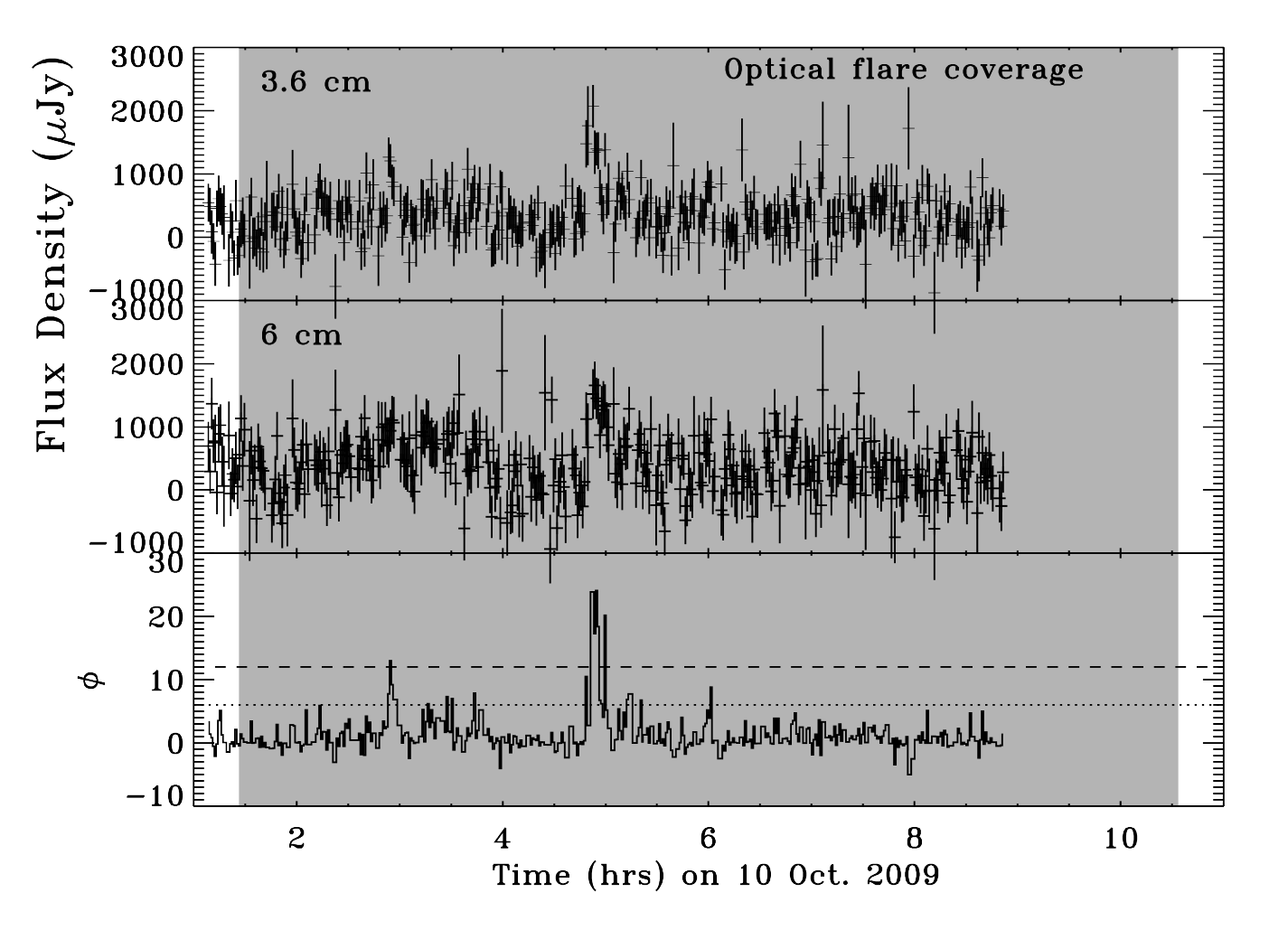}
\caption{Radio light curves. 
 Top panel of each subfigure displays the 3.6 cm light curve and 1$\sigma$ errors, middle panel displays 6 cm light curve and 1$\sigma$ errors, and bottom panel shows value of $\phi$ calculated for each 60s time bin. 
 $\phi$ is a discriminator of correlated multi-band variability, and is defined in Equation~\ref{eqn:phi}. Dotted line shows $\phi=6$ line, and dashed line
indicates $\phi=10$ line.
Grey shaded region indicates the time of optical observation coverage.
\label{fig:rad_opt_coverage}}
\end{figure*}

The observational setup obtained a scan of the flux calibrator to establish the gain solution, then interleaved observations of the target
and the phase calibrator. Calibration and data reduction proceeded using standard techniques for gain and phase calibration for continuum observations of point
sources using AIPS software \citet{aips}. Briefly, the field was imaged once calibration solutions had been obtained and applied to the target source.
The visibilities of all objects in the field that were not EV Lac  were removed from the original dataset to create a dataset containing
visibilities of only the target, EV~Lac. The AIPS task DFTPL applied to these datasets then returned light curves of total intensity and
circularly polarized intensity. Light curves created in time bins of 300 second and 60 second enabled identification of flares and shorter
time binning to investigate the details of flares. There was an issue with the polarization data for the X-band time series on the night of 20 Sept. 2009.
The key quantities of interest for this study are the flux density variation with time, F$_{\nu}$(t),
and the spectral index $\alpha$ between the
two frequencies (determined from assuming F$_{\nu}\propto \nu^{\alpha}$), and its variation with time, $\alpha(t)$.
Additionally the percentage of circular polarization and its variation with time, $\pi_{c}(t)=V(t)/I(t) \times 100$, 
is of interest.
Light curves at the two observing frequencies are shown in Figure~\ref{fig:rad_opt_coverage}.

\begin{figure}[ht!]
\plotone{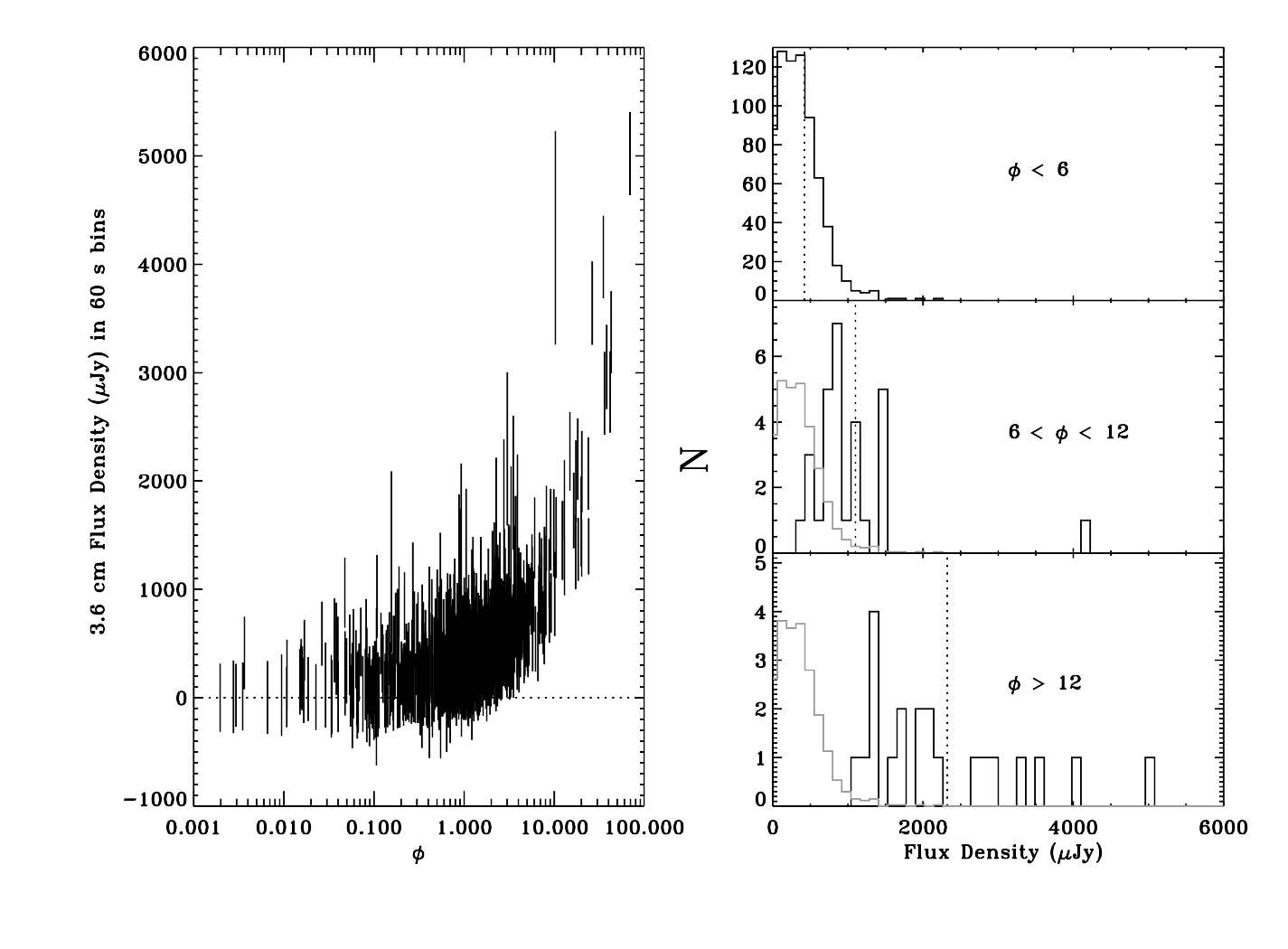}
\caption{
\textit{(Left panel)}
Scatter plot of $\phi$ against 3.6 cm flux density in $\mu$Jy; only positive values are shown.  Each flux density measurement is accompanied by error bars.
\textit{(Right panels)} Distribution of flux densities in several bins of $\phi$.
The vertical dotted line indicates the average flux density in each histogram.  
The distribution for $\phi<6$ is reproduced in the other $\phi$ distribution plots with a renormalization in grey.
While the flux density variations in the uppermost panel indicate noise, and those in the bottom panel are clear measurements of correlated variability, the intermediate case demonstrates situations of correlated variability  at lower S/N.
\label{fig:phixc_vs_x}
}
\end{figure}

\subsection{Optical Observations\label{sec:opt_obs}}

The paper makes use of data from several telescopes at Apache Point Observatory (APO).
As described in K13, U-band data from the New Mexico State University (NMSU) 1 m telescope
were taken with 4 s integrations, with a readout
of about 10 s. 
Every 39 exposures an automatic focus check was performed resulting in a
slightly larger gap in the data.
Observations were reduced as part of the standard
1 m pipeline.
The Flarecam instrument \citep{Hilton2011} on the 0.5 m  ARCSAT telescope at APO was remotely operated during the observations. The SDSS u filter was used on this telescope, with a readout of about 1 s and cadence of 8 s.
The
data were reduced using standard IRAF procedures.

In addition  spectra were obtained with 
the Dual-Imaging Spectrograph
(DIS) on the ARC 3.5 m telescope at APO.
The low-resolution gratings B400/R300 provide 
continuous wavelength coverage from
$\lambda \sim$ 3400–9200 \AA.
From Table~2 and associated discussion in K13, the total exposure time for this night was 8.202 hours, with
exposure times of 6, 15, 20, 30 seconds.
The 1.5'' slit provided resolution at the wavelength of He~I $\lambda$4471 of 5.5–7.3 \AA\ (R $\sim$ 600–800).
We utilize the flare-only spectra as a function of wavelength and time, obtained after scaling the original flux (using the algorithm described in Section 2.6 and Appendix A of K13) and after subtracting an overall quiescent (or pre-flare) spectrum. 
These spectra are suitable for blue continuum analysis.
No short-cadence spectra at red wavelengths ($>$5200 \AA) are available on this night, limiting the available wavelength coverage to 3400 - 5200 \AA.
Data from the entire night of observation on 10 Oct. are contained in extensions 11 and 12 of the file
\verb+r_flux_nB.fit+ available from on-line data of K13, so we use this time-resolved spectroscopic data to examine the weaker flares not individually analyzed in K13.

U band photometry was obtained at the Kitt Peak National Observatory (KPNO) Wisconsin Indiana Yale NOAO (WIYN) 0.9m telescope on the night of 2009 Oct. 8.  
Data were obtained with 15 second exposures, with one observation roughly every 51 seconds. 
Narrow-band H$\alpha$ and H$\alpha$ continuum filter observations 
were obtained
every 6-7 minutes or so.
This data was reduced following standard procedures and is discussed in 
\citet{tofflemire2012}. 
Only 5.5 hours of data were obtained with the telescope due to humid conditions, which necessitated telescope closure.
We note \cite{tofflemire2012}
report on a small flare from EV~Lac on 8 Oct. 2009 starting at about 8.96 hours, which is the just after when the radio observations stopped.

Observations at the Dominion Astronomical Observatory (DAO) obtained data with the 1.8m Plaskett telescope using the SITe5 CCD and Cassegrain Spectrograph, with resolution $\sim$750 spanning  3540-4710 \AA, integration time between 60 and 420 s, and cadence of between 160 and 300 seconds between integrations. 
Further discussion of this setup is given in \S2.4 of K13 as well as \S2.4 of \citet{Schmidt2012}.
Time series of the emission lines of 
H$\gamma$, H$\delta$, Ca~II K on 20 Sept. 2009 only show an appreciable flare starting at 10 hours, right after the radio observations ended.

\section{Data Analysis \label{sec:analysis}}

As described above in \S~\ref{sec:opt_obs} and below in \S~\ref{sec:multiw}, there is no credible evidence of flaring in the data on 20 Sept. or 8 Oct. during the windows where both radio and optical data were  obtained. After a brief description of the entire radio data set in \S~\ref{sec:multiw}, we focus on data from 21 Sept. and 10 Oct. for the rest of this paper.

\subsection{Multi-Frequency Behavior of Large Radio Flares \label{sec:multiw}}

Because incoherent gyrosynchrotron flares emit over a broad bandpass, we expect the flaring variability between the two bands utilized here to be correlated. 
We use a modified Welch-Stetson statistic \citep{Welch1993} to characterize the joint multi-frequency light curve behavior. We denote the statistic  $\phi$, defined as \\
\begin{equation}
\phi = \frac{F_{3.6cm}}{\sigma_{3.6cm}} \frac{F_{6cm}}{\sigma_{6cm}}\;\; , \label{eqn:phi}
\end{equation}
where $F_{3.6cm}$ ($F_{6cm}$) is the flux density in a time bin measured at 3.6 (6) cm, and $\sigma_{3.6cm}$ ($\sigma_{6cm}$) is the uncertainty on the
flux density measurement at 3.6 (6) cm.
This statistic is essentially the product of the signal-to-noise ratios of data taken at the same time, and can be used to investigate signals that might be present at a lower statistical significance in either light curve. 
Figure~\ref{fig:phixc_vs_x} displays a scatter plot of positive $\phi$ statistic values versus the flux density in each 60 s bin. 
There is a wide range of $\phi$ values, 
and a correlation at higher values with the 
flux density.  
The right panels show histograms of the flux density distribution for several cuts of
$\phi$, along with the average flux density value of all the data in that $\phi$ range. 
The data points in the $\phi < 6$ distribution correspond to noise, with the average flux density measured being the same as the median flux density error. 
For $6 < \phi < 12$
the average flux density is several times larger. 
We computed the Student's t-distribution for the $6 < \phi < 12$ and $\phi \ge 12$ distributions against the distribution for $\phi < 6$, with the latter as the control. For 
$6 < \phi < 12$ ($\phi \ge 12$),
the T statistic is -8.7 (-8.5), with probability 3$\times$10$^{-9}$ ($2\times10^{-8}$) of having the same mean as the $\phi < 6$ sample.

Based on this, we  employ a minimum $\phi$
 value of 6 when investigating the radio light curve behavior to identify potential radio flare candidates.
 We can also categorize these as weak or strong flares on the basis not only of the individual flux densities, but also the $\phi$ value at peak.
 For peak $\phi$ values $6 < \phi < 12$ we require multiple successive 60s bins to have $\phi$ values above this threshold
 before considering the variation to be a flare candidate. 
 Regions of the light curve with $\phi >12$ we identify as flare candidates based on only one point above $\phi >12$.
 Figure~\ref{fig:rad_opt_coverage} 
 shows the 3.6 and 6 cm light curves, along with the $\phi$
values, for each of the four days of radio observations.
Lines of $\phi=6$ and $\phi=12$ demarcate regions of no, weak, and strong correlated variability.
From these figures it is clear that there is no evidence for flaring variability on either 20 Sept. nor 8 Oct. 2009.
As described in Section~\ref{sec:opt_obs}, optical flaring on these nights occurred after the radio observations ended. 
There are candidate radio flares on each night of 21 Sept. 
and 10 Oct., which we consider further.

Key characteristics of these  radio flares are tabulated in Table~\ref{tbl:radio}.
For each receiver band the time of peak flux density and the value of peak flux density are listed, along with the exponential decay time from a fit to the decay phase of the flare at each frequency. The peak value of $\phi$ during the flare is also given. 
The polarization information was largely inconclusive due to the large errors; the signal-to-noise ratio of circular polarization measurements for all but the brightest flare are consistent within the large error bars with zero polarization measurement. 
The expectation is that gyrosynchrotron flares should have low levels of circular polarization \citep{dulk1985}, which appears to be confirmed with the flares observed here.

\begin{figure}
    \plotone{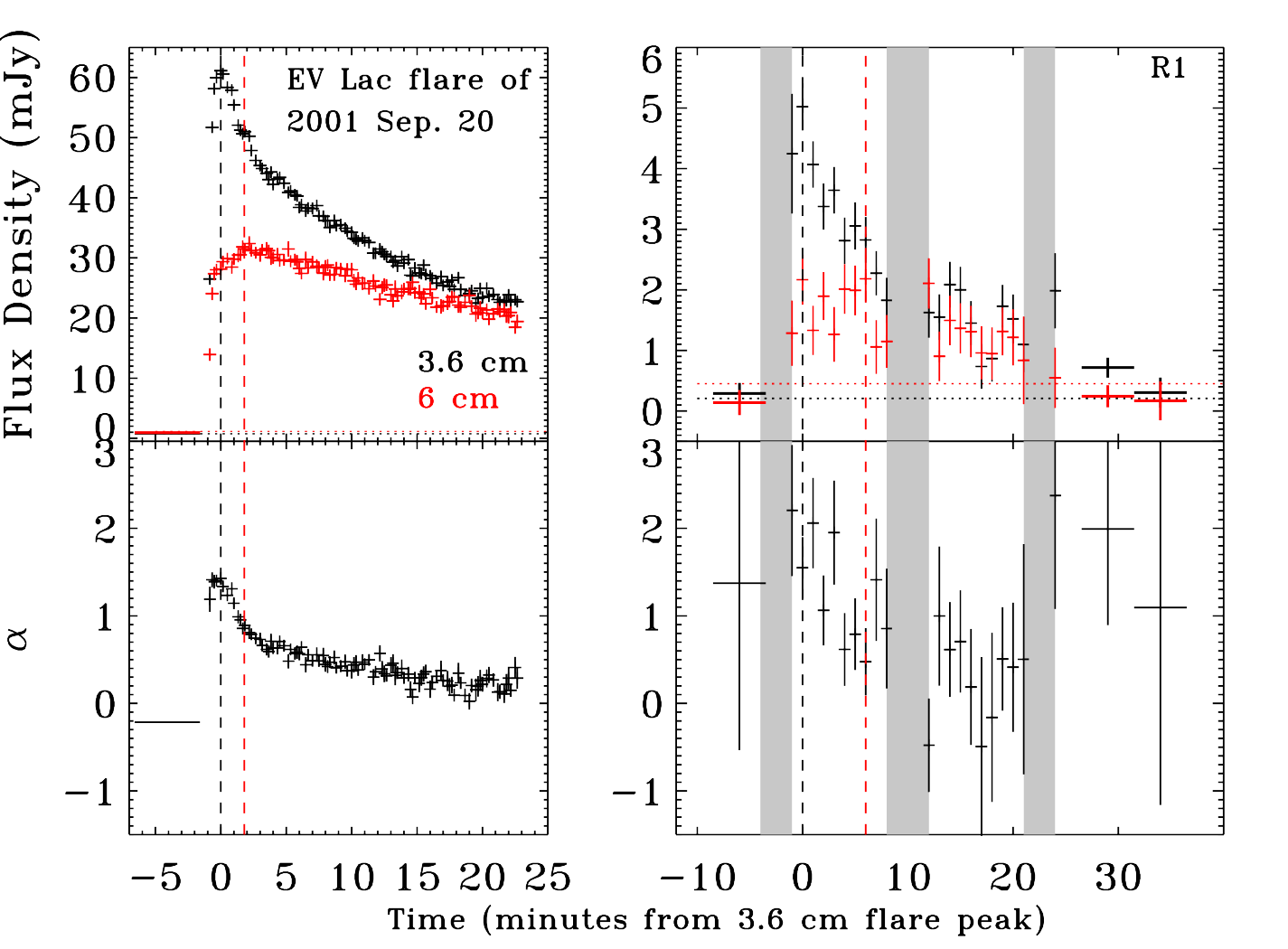}
    \caption{ Light curves and spectral index variations during two remarkable multi-frequency radio flares. Top panels display 3.6 cm (6 cm) band light curves  in black (red), with 1$\sigma$ errors, with flux densities in mJy (=1000 $\mu$Jy). The black dotted line indicates the level of quiescent emission before the start of the flare.
    Bottom panels show variation of spectral index $\alpha$ during the flare. 
    Black dashed line indicates the time of maximum observed flux at 3.6 cm, while red dashed lines indicate the time of 6 cm flux density maximum. 
    During each flare the light curves are shown with time bins of 10s (60s) for the 2001 Sept. 20 (2009 Sept. 21=R1) flare, while before and after a larger time bin of 300s is employed. 
    Grey shaded regions indicate times of calibrator scans.
      The large flare on EV~Lac analyzed in \citet{ostenetal2005} is shown on the left for comparison with the largest radio flare in our sample (R1) on the right.
      Both display a frequency-dependent delay in the time of peak flux density, with the delay increasing to lower frequencies.
    \label{fig:bigradio}
    }
\end{figure}

The largest radio flare observed over the four days in 2009 occurred on 21 Sept. 2009 and did not have overlapping optical data.
Nevertheless, it is intriguing.
The characteristics of this flare are very similar to the 
more than a factor of 10 larger amplitude flare seen on EV Lac on 20 Sept. 2001, reported in \citet{ostenetal2005}, and we show the two
side-by-side in Figure~\ref{fig:bigradio}.
In both cases there is a noticeable delay between the time at which the 3.6 cm light curve reaches its maximum, and the peak time of the flare at 6 cm (which occurs later). 
In the 2001 Sept. 20 flare this difference is about 2 minutes, while for the 2009 Sept. 21 flare it is about 
three times longer. 
In addition to the frequency-dependent delay in the time of peak flare flux, we note that the spectral index 
changes systematically as a function of time between the time of the 3.6 cm flare peak and the time of the 6 cm flare peak in both events.
In both cases the spectral index is positive but declining with time.  
It is clear that the 2001 Sept. 20 flare 
exhibited a spectral index flattening just before the 3.6 cm flux density peak, and the
trend in the spectral index decay changes just after the 6 cm reaches its peak flux density.
There is insufficient signal to noise for the 2009 Sept. 21 flare to determine if similar trends exist. 
In both cases the spectral index was in the range $0 < \alpha <3$, and the decay time was different at the two frequencies (although more similar for R1).

\begin{deluxetable}{lllllll}
\tablewidth{0pt}
\tablecolumns{7}
\tablecaption{Characteristics of radio flares at peak \label{tbl:radio}}
\tablehead{  \colhead{Flare} & \colhead{Date} &  \colhead{Receiver } & \colhead{Time (hr)} & \colhead{Flux Density} & \colhead{$\phi_{\rm peak}$} & \colhead{$\tau_{\rm decay}$}   \\
\colhead{} & \colhead{} & \colhead{Band (cm)} & \colhead{hr} & \colhead{($\mu$Jy)} & \colhead{} & \colhead{(s)} }
\startdata
R1 & 21 Sept. 2009 & 3.6 & 2.59860 & 5021$\pm$384 & 69 & 747.6$\pm$0.4 \\
      &                        & 6 & 2.70418 &  2182$\pm$394 & & 1035$\pm$133  \\
R2 & 21 Sept. 2009 & 3.6 & 7.89862 &2202$\pm$378 & 18.2 & 228$\pm$96\\
   &               & 6    & 7.93750 &1592$\pm$694 &  & 141$\pm$74\\
   R3 & 21 Sept. 2009 & 3.6 &8.34862 &1705$\pm$512 &8.7 & 612$\pm$144\\
       &               & 6  &8.33194 &1058$\pm$430&  & 324$\pm$216 \\
R4 & 10 Oct. 2009 & 3.6 &2.89306 &1266$\pm$310 &13 &266$\pm$82 \\
    &              & 6 &2.92639  & 944$\pm$384 & & 418$\pm$166 \\
R5 & 10 Oct. 2009 & 3.6 &4.87639  &2070$\pm$337 & 24 & 557$\pm$81\\
     &            &      & 4.89305& 1402$\pm$377& &732$\pm$143 \\
\enddata
\end{deluxetable}

We can examine some properties of the radio-emitting flares, under the assumption that the radio-emitting particles experience strong diffusion,
as observed in solar radio flares \citep{Lee2002}, and likely due to increased turbulence from wave-particle interactions 
\citep{MelroseBrown1976}.
The so-called ``trap-plus-precipitation'' model
for energetic electrons in solar flares bifurcates the action of accelerated particles depending on their pitch angle relative to the magnetic field line. Once particles have been injected into a closed magnetic loop,
those which have a pitch angle less than a critical value will immediately precipitate or leave the loop, depositing their energy in the lower atmosphere. Particles having a loss cone pitch angle larger than this critical value will be
mirrored in the coronal loop, since a converging magnetic field acts as a natural magnetic trap. 
The slow decay implies long-term trapping and a consequent large mirror ratio \citep{Lee2002}.
We follow the reasoning in \cite{Lee2002} identifying the decay time of the radio flare as indicative of the precipitation rate in a trap plus precipitation model, where at late times the
radio flux density proceeds as $F(t)\propto e^{-\nu t}$, with $\nu$ the precipitation rate. 
The critical pitch angle $\theta_{0}$ can be expressed as the ratio between the magnetic field strengths at the foot points ($B_{0}$) and loop top ($B_{1}$)
of a magnetic structure, according to \\
\begin{equation}
  \sin \theta_{0}=\left( \frac{B_{1}}{B_{0}} \right)^{1/2} \;\;\;
\end{equation}
which in the small angle limit reduces to $\theta_{0}=\left( \frac{B_{1}}{B_{0}} \right)^{1/2}$.
The precipitation rate $\nu$ can be written in the strong diffusion limit
as \\
\begin{equation}
    \nu = \frac{1}{2} \theta_{0}^{2} \frac{c}{H_{1}-H_{0}}
\end{equation}
where the structure extends from $H_{0}$ (with associated field strength $B_{0}$) to $H_{1}$ (with associated field strength $B_{1}$)
above the surface of the star \citep{2006A&A...453..959M}. 
For a dipole configuration, $B_{1}/B_{0}$=$\left( \frac{H_{0}}{H_{1}} \right)^{3}$.
To match the observed exponential decay of the flux density light curve, the precipitation rate will be equal to the inverse of the decay timescale, giving 
$\nu = 1/\tau{_d}$. 
This results in a relationship between the decay timescale of the flare and the two distances, \\
\begin{equation}
    \frac{1}{\tau_{d}}= \frac{1}{2} \left( \frac{H_{0}}{H_{1}} \right)^{3} \frac{c}{H_{1}-H_{0}} \;\;\; .
\end{equation}
This can be rearranged to give a function f($H_{0},H_{1}$)\\
\begin{equation}
   f(H_{0},H_{1}) = \frac{2 (H_{1}-H_{0})}{c \tau_{d}} -\left( \frac{H_{0}}{H_{1}} \right)^{3}
\end{equation}
and zeroes of the function, and thus appropriate values of $H_{1}$, can be determined once
$H_{0}$ is specified. 
For simplicity we assume $H_{0}=1$, i.e., the loop's footpoint is at the stellar surface, and express the 
loop height $H_{1}$ relative to 
the stellar surface, i.e. plotting 
$H_{1} (R_\star)-R_\star$. 

\begin{figure}
    \includegraphics[scale=0.6]{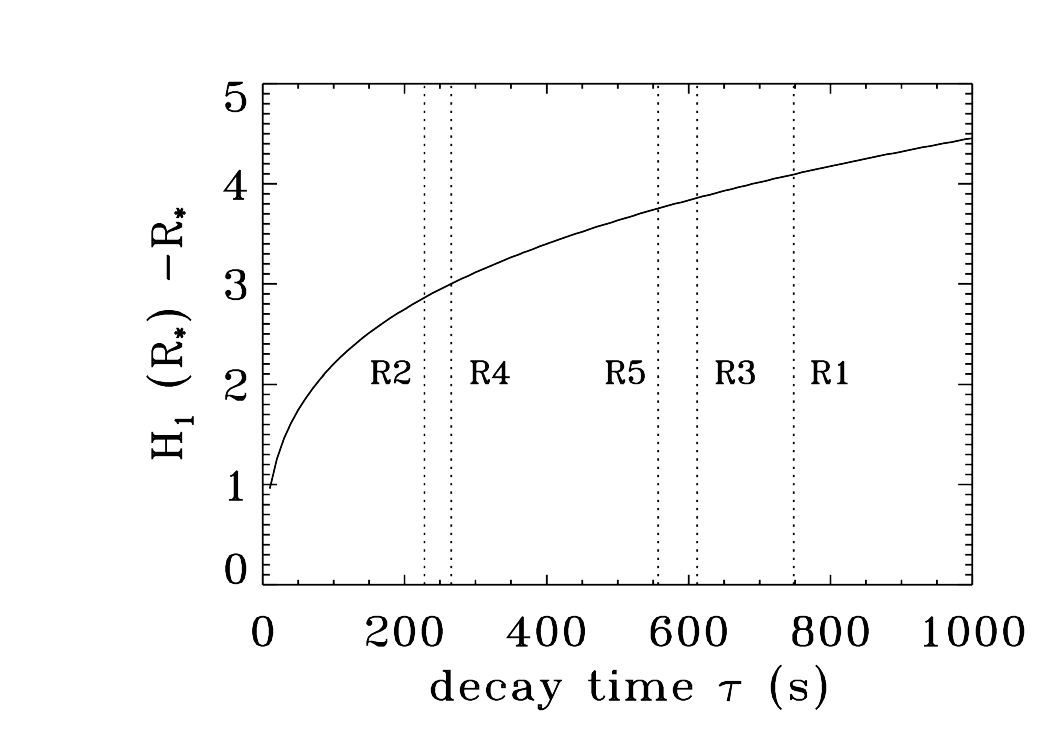}
    \caption{Constraints on loop height as a function of observed radio flare decay time, under assumptions of strong diffusion and a single loop.
    Dotted lines show decay time of the radio flares considered here (taken from Table~\ref{tbl:radio}).
    \label{fig:loop_decay}}
\end{figure}

Figure~\ref{fig:loop_decay} shows the height of loops above the surface which start at the stellar surface, as a function of the radio decay time $\tau_d$; values for the flares under consideration in this paper are taken from Table~\ref{tbl:radio} and also indicated in Figure~\ref{fig:loop_decay}.
These loops appear to be fairly extended, with heights ranging from 3-4 R$_{\star}$ above the surface.
Making an assumption that the loops are approximately semi-circular shape, an upper limit to the loop length can be obtained with L$\approx \pi$H$_{1}$, giving loop lengths of 9-12 R$_{\star}$.

\citet{benz2002} describes the condition for strong diffusion as the diffusion timescale $\tau_{d}$ being less than the lifetime of a particle in the magnetic trap. 
As noted above, wave-particle interactions can enhance the diffusion timescale above what would be expected based only on collisional processes.  We can estimate the maximum escape time for a particle in the loss cone at the loop top and at the largest pitch angle within the loss cone, $\tau_{e}$,
\begin{equation}
    \tau_{e}=\frac{\pi}{4} \frac{L\sqrt{M}}{v_{z}^{\rm top}}
\end{equation}
with $L$ the loop length, $M$ the magnetic mirror ratio
$=B_{0}/B_{1}$, and $v_{z}^{\rm top}=v \cos \theta_{0}$
given the particle velocity $v$ and critical loss cone 
angle $\theta_{0}$. 
The total lifetime is given by $\tau_{e} M$ \citep{benz2002}.
For the results above, the magnetic mirror ratio ranges from 60-130, with maximum escape time for a 30 keV trapped electron ranging from 2.2-10.6 hr, and for a 1 MeV trapped electron from 0.8-3.7 hr. These are comfortably larger than the observed flare timescales of several minutes, supporting the 
initial assumption of strong diffusion.
If we further assume that collisions dominate the diffusion, then we can equate $\tau_{d}$ with  the collisional deflection time $\tau_{\rm defl}$ and place an upper limit on the ambient electron density in the radio-emitting region, for a representative 1 Mev electron.
In this case 
\begin{equation}
\tau_{\rm defl} = 0.95\times10^{8} \left( \frac{E_{\rm keV}^{3/2}}{n_{e}}\frac{20}{\log \Omega} \right)\;\; ,    
\end{equation}
with $E_{\rm kev}$ the electron energy in keV, electron density $n_{e}$, and Coulomb logarithm $\log \Omega=\log (8\times10^{6}T/\sqrt{n_{e}})$\citep{benz2002}. 
Given the timescales for these radio flares in Table~\ref{tbl:radio}, the implied electron densities $n_{e}$ are all less than a few times 10$^{8}$
cm$^{-3}$, with the shortest events (R2 and R4) compatible with $n_{e} < 10^{9}$
cm$^{-3}$.
If wave-particle interactions provide an additional source of diffusion then $\tau_d$ could be less than this value, so the ambient density could be larger that these limits and still be consistent.

\subsection{Co-Occurrence of Radio and Optical Flares \label{sec:rad+opt}}

\begin{figure}
    \plotone{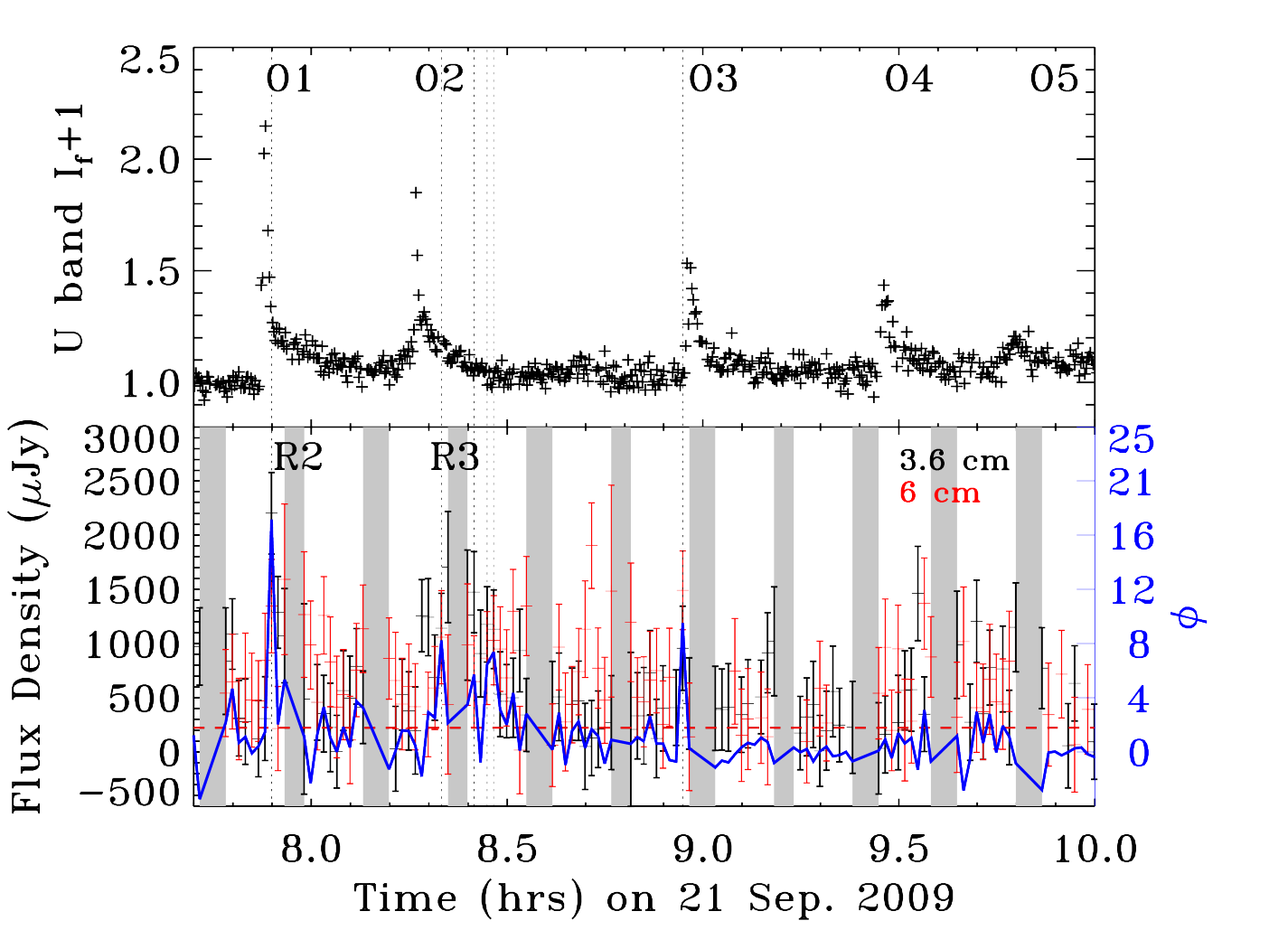}
    \caption{
    Overlap of optical photometry and radio light curves on the night of 21 Sept. 2009, during times of optical flare activity. 
    Specific optical flares are labeled in the top plot.
    Top panel displays optical photometric brightness variations, while bottom panels display 3.6 cm (6 cm)  light curves in black (red), along with the $\phi$ parameter in blue, as discussed in the text. Dotted lines connect time bins in the radio light curve with $\phi > 6$ with the optical light curve.  
    The quiescent flux density values at 3.6 (6) cm are indicated in dotted lines in black (red), respectively.
    Light gray regions indicate gaps in data stream due to phase calibrator observations.
    \label{fig:sep21_rad+opt}
    }
\end{figure}

\begin{figure}
    \plotone{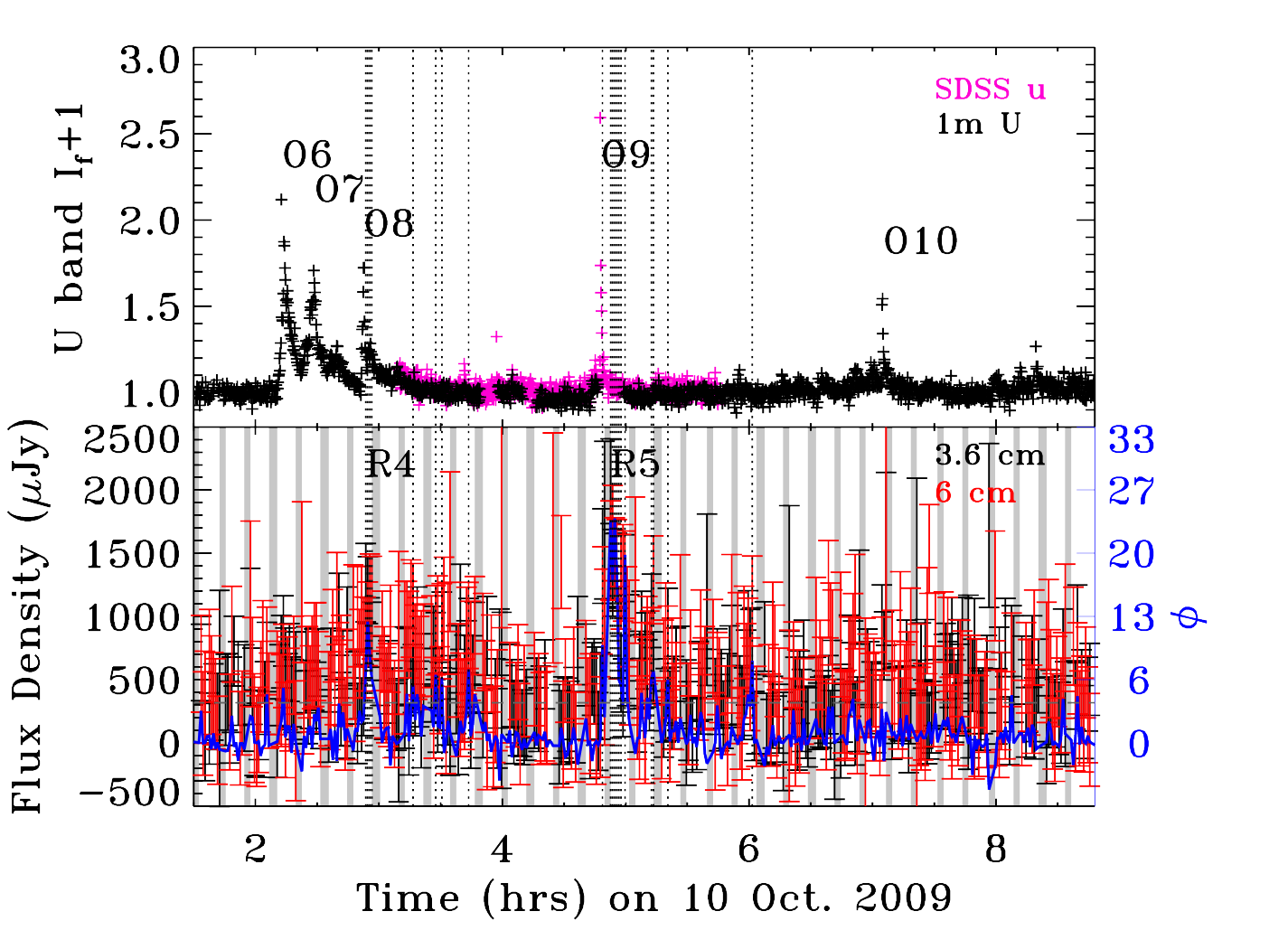}
    \caption{
   Same as Figure~\ref{fig:sep21_rad+opt}, but for data on 10 Oct. 2009. 
    \label{fig:oct10_rad+opt}
    }
\end{figure}

 Figures~\ref{fig:sep21_rad+opt} and ~\ref{fig:oct10_rad+opt}
 display the optical and radio light curves for 21 Sept. and 10 Oct. respectively. On 21 Sept. there was a sequence of $\approx$ 5 optical flares with sequentially decreasing peak amplitudes.
 On the night of 10 Oct. there is again a sequence of multiple optical flares -- a complex of 3 closely occurring flares between 2 and 3 hours, and two successive flares around 5 and 7 hours, respectively.  
We use the photometric data to determine several quantities about the flares, which are tabulated in Table~\ref{tbl:opt}. 
The peak time is the time of maximum flare amplitude. 
The quantity $I_{f}+1$ is the flux enhancement relative to the quiescent flux; $t_{1/2}$ is the full-width at half maximum of the light curve. 
This quantity is determined by resampling the light curve at a factor of 10 times higher time resolution, then determining the width at half maximum. 
The flare impulsiveness index $\mathcal{I}$ is defined as  
$\mathcal{I}= I_{f,peak}/t_{1/2}$, a measure of the peak amplitude of the flare relative to its timescale.
This index is described in more detail in K13.

The integrated energy in the U band is also tabulated; this was calculated by determining the equivalent duration, the integral of $I_{f}$ over the duration of the flare, multiplied by the quiescent U band luminosity of EV~Lac (given in K13 as 
10$^{28.8}$ erg s$^{-1}$). 
Using the nomenclature of K13 for impulsive, hybrid, and gradual flares ($\mathcal{I}>1.8$, 0.6$<\mathcal{I}<$1.8, and $0.6 <\mathcal{I}$),
respectively, it is clear that only one of these flares would be an impulsive event (O9), with six gradual events and three hybrid events. 
We note that flare O7 in the present work has been analyzed by K13  as event GF3,
and flare O9 analzed as event IF6.

\begin{deluxetable}{lllllll}
\tablecolumns{7}
\tablecaption{Characteristics of optical flares \label{tbl:opt}}
\tablehead{  \colhead{Flare ID} & \colhead{Peak Time (hr) } &  \colhead{I$_{f,peak}$+1 } & \colhead{t$_{1/2}$} & \colhead{$\mathcal{I}$} &  \colhead{E$_{U}$}  &\colhead{$\chi_{\rm fl,pk}$}\\
\colhead{} & \colhead{} & \colhead{} & \colhead{(min)} & \colhead{} & \colhead{(10$^{30}$ erg)} & \colhead{} }
\startdata
O1 & 21 Sept. 7.88309 & 2.15 & 0.87 & 1.3 & 11.5 & \ldots\tablenotemark{s} \\
O2 & 21 Sept. 8.26706 & 1.85 & 0.56 & 1.5 & 8.8 & \ldots\tablenotemark{s} \\
O3 & 21 Sept. 8.9599 & 1.533 & 1.67 & 0.3 & 7.7 & \ldots\tablenotemark{s} \\
O4 & 21 Sept. 9.46195 & 1.435 & 4.11 & 0.1 & 5.7 &\ldots\tablenotemark{s} \\
O5 & 21 Sept. 9.83105 & 1.23 & 7.02 & 0.03 & 4.3 & \ldots\tablenotemark{s} \\
O6 & 10 Oct. 2.21047 & 2.12 & 3.13 & 0.4 & 15 & 2.6\\
O7\tablenotemark{a} & 10 Oct. 2.4744 & 1.87 & 2.88  & 0.3 & 35 & 2.99$\pm$0.31\\
O8 & 10 Oct. 2.87686 & 1.72 & 1.18 & 0.6 & 10 & 2.7\\
O9\tablenotemark{b} & 10 Oct. 4.7934 & 2.6 & 0.32 & 5 & 4 & 2.19$\pm$0.20\\
O10 & 10 Oct. 7.08125 & 1.54 & 0.67 & 0.8 & 2.6 & 8.1\\
\enddata
\tablenotetext{a}{This flare is GF3 in the study of K13; parameters taken from Tables~6 and 7 of that paper.} 
\tablenotetext{b}{This flare is IF6 in the study of K13; parameters taken from Tables~6 and 7 of that paper.}
\tablenotetext{s}{These flares lack accompanying spectroscopic data.}
\end{deluxetable}

\begin{figure}
    \plotone{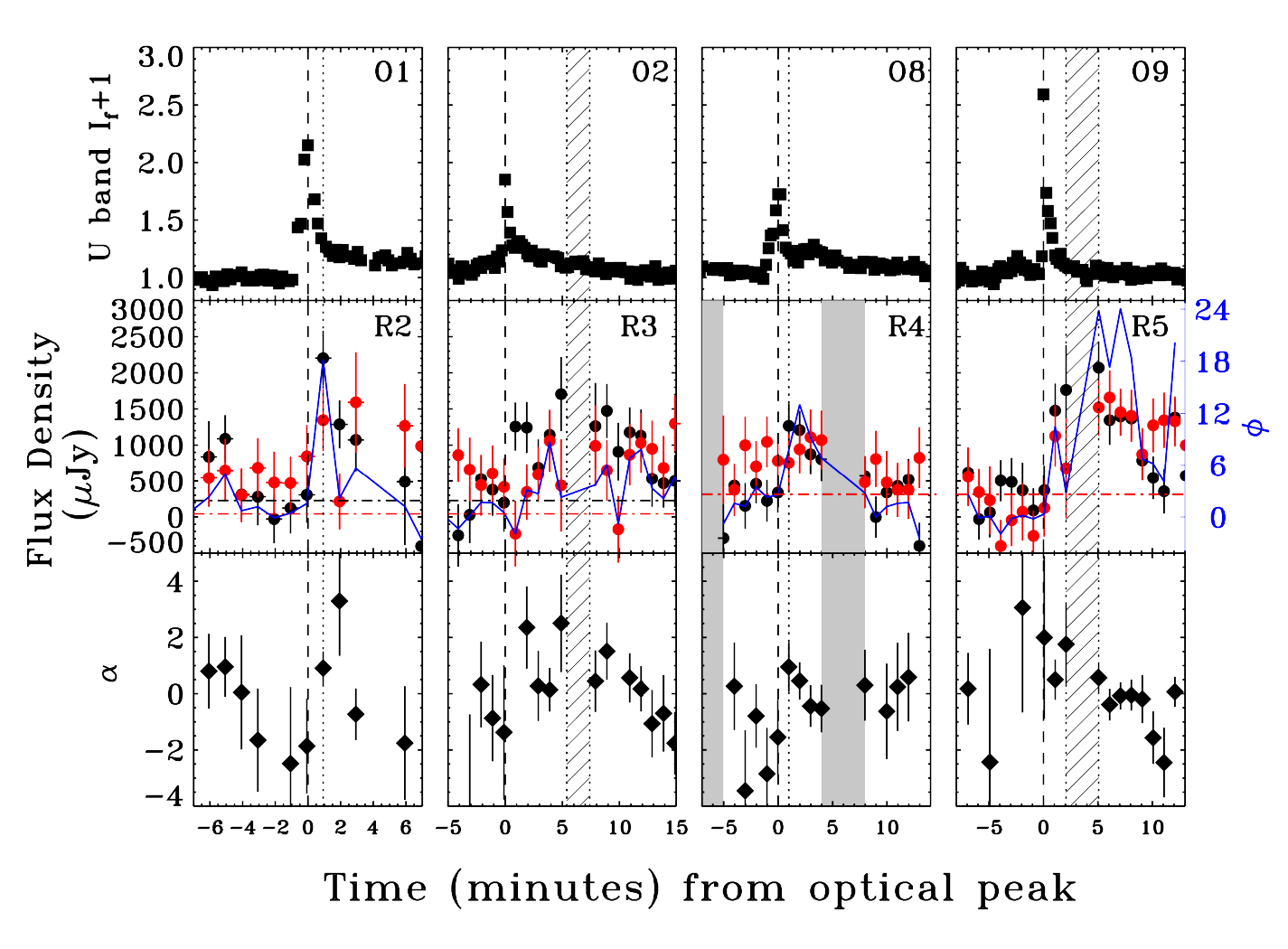}
    \caption{Close-up of four flares with radio and optical response.  Top panel plots optical photometry around the time of the optical peak, middle panel displays 3.6 cm (6 cm) radio light curve in black (red), and bottom panel displays the spectral index $\alpha$. Tables~\ref{tbl:radio} and ~\ref{tbl:opt}   list the radio and optical properties of the flares, respectively.  Dashed lines indicate the time of the optical flare peak, while dotted lines indicate the time of the 3.6 cm flux density peak. 
    Light grey shaded regions indicate times of phase calibrator scans.
   Dash-dot lines in black (red) show the quiescent flux density level at 3.6 (6) cm, respectively, on that day.
   For R3 and R5, a calibrator scan indicated by the hatched shading likely hides the time of true flux density peak. 
\label{fig:rad+opt_flares}}
\end{figure}

The radio-optical joint light curves in Figures~\ref{fig:sep21_rad+opt} and \ref{fig:oct10_rad+opt}
 reveal an inconsistent pattern of multi-wavelength responses.
The data suggest an overall low correspondence between production of optical and radio flares. 
From these data we see that out of 10 optical flares (9 with suitable radio coverage) there are only 4 with an appreciable radio response, three of which exhibit a strong response ($\phi >12$). 
Figure~\ref{fig:rad+opt_flares} shows a close-up of the times around these events. 
Intriguingly, there is a common characteristic for all events with a radio and optical response that the optical light peaks before the radio wavelength emission reaches its maximum.
This delay is of order 1-2 minutes for the cases of O1/R2 and O8/R3. The events O2/R3 and O9/R5 have a gap in the radio data due to a calibration scan, and it is not clear at what time the peak flux density occurs. 
The delay appears to be bounded by 5-7 minutes in the case of O2/R3 and 2-5 minutes in the case of O9/R5. 
Note that \cite{ostenetal2005} also reported a similar phenomena, namely that the 2001 Sept. 20 flare had an optical peak 54 seconds before the 3.6 cm peak.

\subsection{Time-resolved Optical Spectroscopy \label{sec:spec+time}}

\begin{figure}
    \includegraphics[scale=0.5]{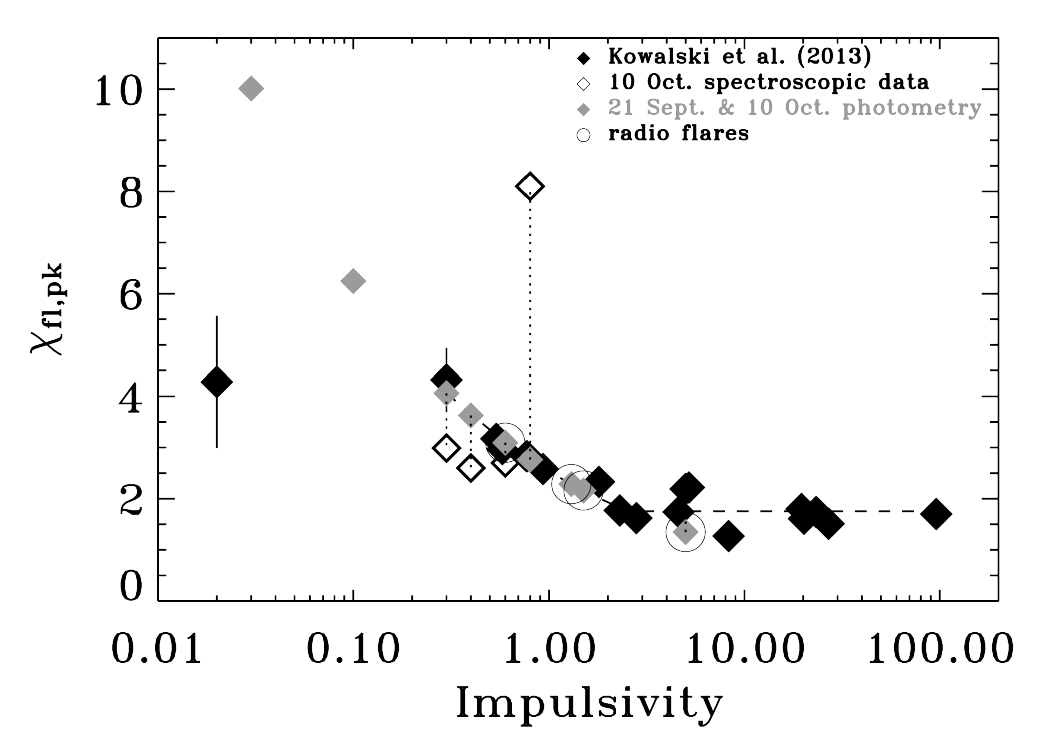}
    \caption{Plot of flare impulsivity versus Balmer jump ratio $\chi_{\rm flare,pk}$
    using the data from K13  in solid diamonds.  Open diamonds are spectroscopic data for additional flares on the night of 10 Oct. which do not appear in the flare spectral atlas of K13.
    Grey diamonds indicate values of $\chi_{\rm flare,pk}$ predicted from a fit to the 
    trend of $\chi_{\rm flare,pk}$ vs. impulsivity using the data in K13. Dotted lines connect the predicted values with flares where
    $\chi_{\rm flare,pk}$ was measured from spectroscopy.  
    Open circles indicate flares from the present study which had accompanying radio flares.
    \label{fig:opt_phot}}
\end{figure}

 K13 established that the Balmer jump ratio $\chi_{\rm flare,pk}$ shows a global trend with the impulsiveness index of the flare.
This quantity $\chi_{\rm flare,pk}$
is defined as the flux ratio from flare-only spectra at the peak of the flare, in two continuum windows between 3600-3630 and 4155-4185 \AA. 
We explore how well the photometric flare properties such as impulsiveness index can predict spectroscopic properties like the Balmer jump ratio in our data. 
We start with the data presented in K13 Table~6 \& 7 to reproduce their Figure~10 ($\chi_{\rm flare,pk}$ versus $\mathcal{I}$), 
then employ the time-resolved spectra from online data noted above to add in $\chi_{\rm flare,pk}$ values for flares O6, O8, and O10 on 10 Oct.  
These numbers are tabulated in Table~\ref{tbl:opt}
and shown in Figure~\ref{fig:opt_phot}.
Below a flare impulsivity value of $\approx$ three,  a trend of increasing Balmer jump with decreasing impulsivity is evident.  
With the spectroscopic data from the flare atlas of K13, we determine a fit to the trend of $\chi_{\rm flare,pk}$
versus flare impulsivity.
Comparison of these predicted values with those measured spectroscopically on 10 Oct. shows a large spread. 
We note that error bars on the time-resolved spectra are not
provided in the on-line  data for the flare spectral atlas, and
the event with $\chi_{\rm flare,pk} \sim$
8, O10, apparently suffers from low signal-to-noise ratio, as the flare-only spectrum at peak dips to below zero past 4400 \AA, and the 4155-4185 \AA\ flux average is only $\approx10^{-15}$ erg cm$^{-2}$ s$^{-1}$ \AA$^{-1}$. 

We then predict what the corresponding Balmer jump ratio would be for the optical flares observed on 21 Sept., using the photometrically measured impulsivity index, as well as the 10 Oct. data to which we compare the measured Balmer jump ratio.
These are shown in grey in Fig.~\ref{fig:opt_phot}. 
The two events with the lowest flare impulsivity, O4 and O5, are  out of the range where there is sufficient data to constrain any relationship, and these are likely unphysical.  The large discrepancy between the measured value of $\chi_{\rm flare,pk}$ and the photometrically-predicted value for O10 likely reflects the low signal-to-noise of the spectroscopic data as noted above.
These results cast doubt on the utility of using photometric data alone to predict the spectroscopic flare properties. 
We note that \citet{Kowalski2016} also reported a variety of Balmer jump ratios for a given flare impulsivity.

\begin{figure}
    \plotone{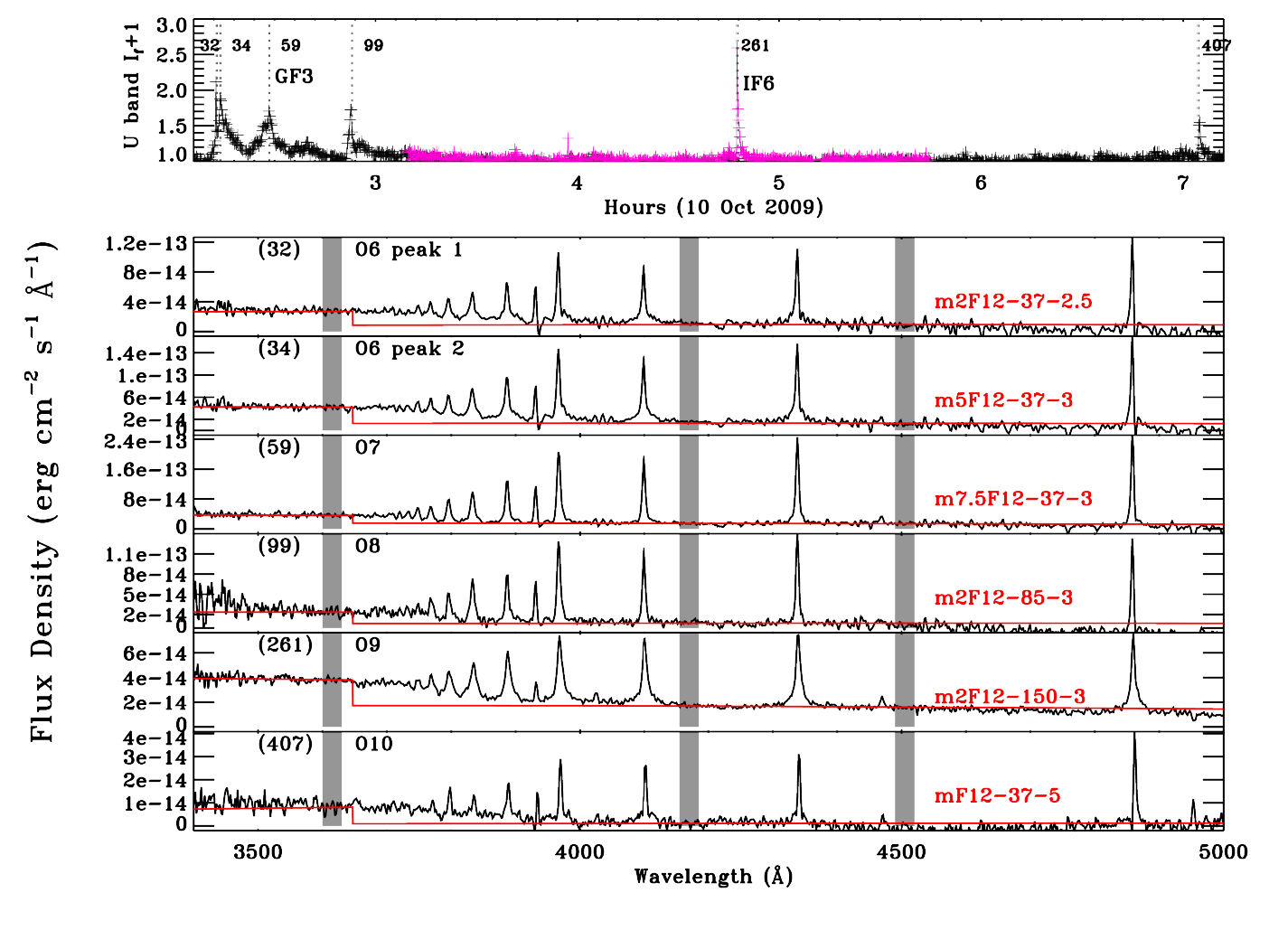}
    \caption{Top panel shows optical photometry  during the time of 10 Oct. 2009. Black curve 
    shows U band from the 1m Telescope; pink curve shows u band from 0.5m telescope. At the peak of each flare the index of the corresponding low resolution optical spectrum number is indicated, and the spectrum is plotted below. 
    The identification of two flares discussed in K13 are also given.
    The lower 6 panels display the optical spectrum from 3400-5000 \AA\ at flare peak.  Grey shaded regions indicate windows used for continuum measurements. Red lines show the best-fit model spectrum to the underlying continuum, and the model name is shown in red. See text for more details. 
    \label{fig:optical_modelfits}
    }
\end{figure}

We next make use of the time-resolved optical spectroscopy from K13 to explore the spectral behavior of the flares on 10 Oct. 2009.
\citet{KAC2024} discusses the utility of using the Balmer jump ratio together with the blue-to-red optical continuum ratio (4155-4185 \AA\ to 5990-6030 \AA) to diagnose the low energy cutoff 
for the accelerated electron distribution used as input to the radiative hydrodynamic models explaining the flare spectral energy distribution. 
Wavelength coverage for this night does not extend redder than 5200 \AA, precluding use of this particular observational diagnostic.
Figure~\ref{fig:optical_modelfits} reproduces the optical light curve, as well as displaying the spectrum closest in time to the peak of the optical flare. 
The times of these spectra are indicated in the light curve along with the index number from the 
K13 on-line data. 
The bottom panels display the pre-flare subtracted spectra from 3400-5000 \AA\ in black. 
All of the spectra exhibit the classic flare continuum increase to blue wavelengths, with a notable increase to the blue towards the Balmer jump, and prominent Hydrogen Balmer emission lines. 

The models in 
\citet{KAC2024} describe the impact to the lower stellar atmosphere from a distribution of energetic electrons, where some parameters of the distribution are varied, as well as the time profile of the injected beam flux. 
A quiescent M dwarf atmosphere is used as initial input.
Some of these parameters are $\delta$, the power-law index of the accelerated electrons above a low-energy cutoff of $E_{c}$,  the total beam flux $F$, and
the description of the injected beam flux time dependence, either constant over 2.3 $s$, or a ramped injected beam flux to the maximum $F$ over that same timescale. 
Tabulated spectral energy distributions are available from their Zenodo repository \citep{KAC2024data}  for a model grid, with the main components ranging over $\delta$ from 2.5-4, $E_{c}$ from 17-500 keV, $F$ from 10$^{10}$-10$^{13}$ erg cm$^{-2}$ s$^{-1}$,
and the two types of injection. 
For the spectrum closest in time to the peak of each flare we determine the best-fit model making use of the tabulated time-averaged spectrum for that model, and averaging over continuum windows from 3600-3630, 4155-4185, and 4490-4520 \AA\ in both data and model.
We use the models which have a ramped injection, $m$, as there were significantly more of these (43 compared to 28 for the constant beam injection). 
Figure~\ref{fig:optical_modelfits} shows the observed spectrum 
of  blue-optical light at the peak time of each flare, as well as the best-fit model.
The name of the best-fit model is also indicated: 
the flare model nomenclature gives first the type of beam injection (=$m$ here as noted above), next the peak beam flux denoted $n$F$x$ for a beam flux of $n\times F^{x}$ erg cm$^{-2}$ s$^{-1}$, the low energy cutoff prepended by a "-", and the power-law index of the distribution prepended by a "-".

The results indicate that moderately large beam fluxes greater than or equal to 10$^{12}$ erg cm$^{-2}$ s$^{-1}$
are required to fit the flare spectral energy distribution at peak for all of these flares. 
In addition, with the exception of the last spectrum which is fit with a larger value of $\delta$ than the others, the electron distribution is fairly shallow and has relatively more high energy electrons compared to low energy electrons (a so-called ``hard'' distribution).
The exception is the last spectrum, but as noted above it is affected by the low signal-to-noise.
The low-energy cutoff for the best-fit model for four spectra is 37 keV, a number commonly assumed in solar flare models \citep{allred2005,holman2003}.  
Two flares have spectra at peak which are best constrained with a higher value of the low-energy cutoff, with O8 (at time index =99) having an $E_{c}$
of 85 keV, and O9 (at time index=261) having an $E_{c}$
of 150 keV.

\section{Discussion \label{sec:disc}}

There are three main items for discussion with this dataset: for flares with both an optical and a radio response, the observed delay between the time of peak flux density; the frequency-dependent delay in radio flares; and  the properties of optical flares occurring with and without a radio response. Each is discussed further in its own subsection below.

\subsection{Optical Flares with Radio Flares: Delay Between Optical Flare Peak and Radio Flare Peak \label{disc:opt+radio}}

The four examples of optical flares with co-occurring radio flares all have the characteristic of a delay between the time of peak emission between the optical and radio flare.
This delay varies from $\approx$ 1 minute to as much at 7 minutes and is in the sense of optical flare peaking before the radio flare.
We consider three different scenarios below to explain this.

The first possibility would be a delay in the arrival time of electrons based on their intrinsic energy. 
On the Sun, time-of-flight differences in arrival times of hard X-ray photons during flares are attributed to the dispersion in electron speed \citep{aschwanden2002} occurring over timescales $\ll$ 1 s.
In this scenario, the more energetic electrons are travelling faster and, assuming they are accelerated at the same time and place, should arrive at the location in the lower atmosphere where they are stopped before the lower energy electrons.
While recent beam heating models \citep{kowalski2015} have shown that energetic electrons can produce the white light flare signature, these have energies above a few tens of keV, with the most dramatic having energies above about 500 keV.  
Radio observations generally probe much more energetic electrons, typically with hundreds of keV to MeV-level energies.  
Not only would this would produce delays in the opposite sense to what is seen, 
the magnitude of the observed timing differences is not compatible with such a scenario, given the timescales of 1-7 minutes noted in Section~\ref{sec:rad+opt}. 
Particle energies in the range 30 keV -2 MeV and plausible loop lengths of $\sim$10 $R_{\star}$ or less (as indicated in \S~\ref{sec:multiw}) would result in arrival time differences of less than $\approx 2$ second.
In order to be consistent with what is observed here, the optical flare would need to represent more energetic electrons than what is producing the radio flare. 
A potential solution for this is if the high energy electrons are accelerated later in the flare.
\citet{Bai1976} determined that in one solar flare the relativistic electrons and energetic nuclei had different acceleration mechanisms from those electrons with energies less than 100 keV, which would be potentially consistent with what is seen here. 
More recently \citet{Arnold2021} simulated electron acceleration during two-dimensional macroscale magnetic reconnection, and found that the high energy tail of nonthermal electrons developed over 5-10 Alfv\'{e}n timescales. Given a magnetic field strength of $\sim$100 G, density of 5$\times$10$^{9}$ cm$^{-3}$ and length scale of $\sim$10$^{9}$ cm, 5-10 Alfv\'{e}n timescales would be of order 15-30 seconds, and so  could be consistent with
the present results.


\begin{figure}
    \plotone{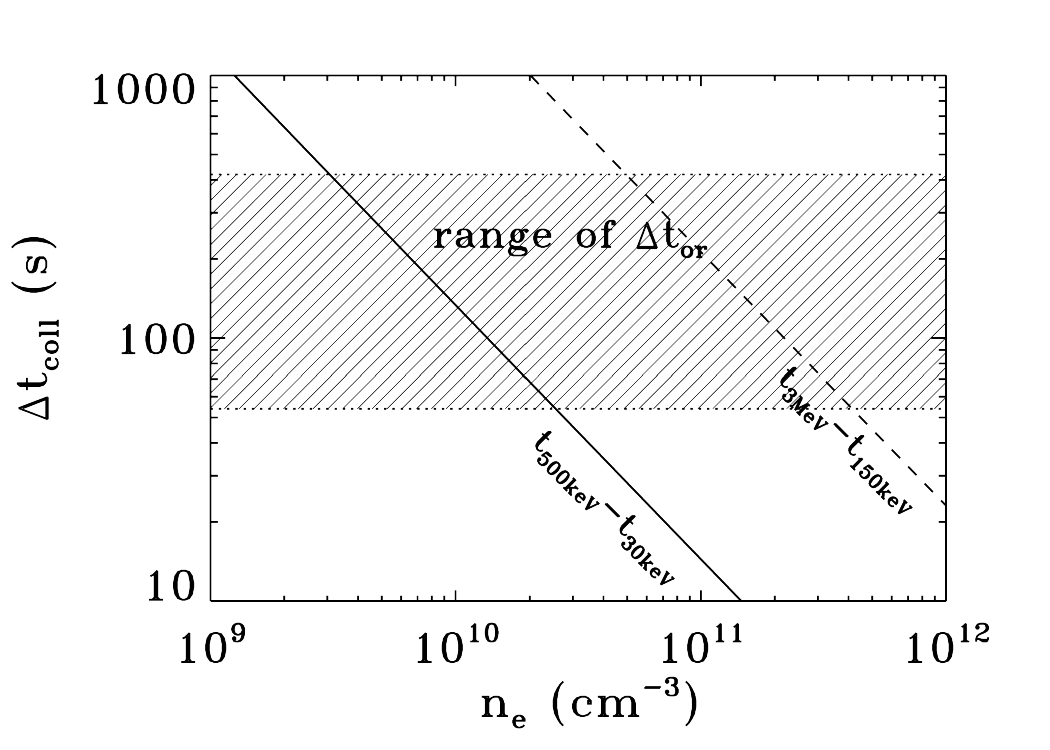}
    \caption{Difference in collisional deflection timescale as a function of electron density, for two sets of electron energies representing likely energies of optical- and radio-emitting electrons, respectively.  Hatched region indicates the range of offsets observed between optical and radio peaks as deduced from Figure~\ref{fig:rad+opt_flares} and discussion in \S~\ref{disc:opt+radio}. This assumes that the difference between the optical and radio flare peaks is due to the trapping time, which will be longer for more energetic particles. \label{fig:tau_defl}}
\end{figure}

The second possibility to explain the time difference is a trapping timescale.  
If the optical flare is produced as a result of directly precipitating electrons of lower energy, they will leave the magnetic trap, interact with the lower atmosphere to produce the optical emission, and 
be gone from the system. In contrast, the radio-emitting electrons can continue to produce radio emission while they are in the trap.  
The radio-emitting electrons will also precipitate from the trap, both directly  and a later time when their pitch angles fall below the critical pitch angle.
If diffusion of particles in the loss cone is dominated by collisions, 
one would expect the trapping time for lower energy electrons to be much shorter than for higher energy electrons, according to the collisional deflection time.
We expect the electrons producing the optical flare to have lower energies, E$\sim$100 keV, and be in a region of higher electron density, $n_{e}\sim$10$^{12}$ cm$^{-3}$, more in line with expectations for weak diffusion 
as inferred from hard X-ray studies of solar flares
\citep{aschwanden2002}, while the electrons producing the radio emission will have higher energies and lower ambient electron densities. 
Both of these factors work in the same way, to decrease the trapping time for electrons producing the white light flare, and to increase the trapping time for electrons producing the radio flare. 
In reality, the electrons producing the white light flare will also experience some trapping, but this timescale is expected to be much smaller than what is seen at radio wavelengths. 
Figure~\ref{fig:tau_defl} shows the timescales for electrons involved in collisional deflection, for two different pairs of electron energies which reflect the range of possible energies associated with optical- and radio-emitting electrons, assuming they occur in the same density region.
The range of timescales corresponding to the observed delay between optical and radio flare peaks is also shown; these are taken from Figure~\ref{disc:opt+radio} and \citet{ostenetal2005}. 
The observed timescales  imply a range from 
$3\times10^{9} \;\;cm^{-3}< n_{e} < 2.5\times10^{10}\;\;cm^{-3}$ 
for 500 keV and 30 keV electrons producing the radio- and optical- flare emissions, respectively, or 
$5\times10^{10} \;\;cm^{-3} < n_{e} < 4\times10^{11}\;\;cm^{-3}$ 
for 3 MeV and 150 keV electrons producing respectively, radio- and optical-flare emissions.
Given the inverse dependence of the collisional deflection time on electron density, correcting for the higher density of the optical flare-emitting region compared to the radio flare-emitting region will be a minor effect. 

The third possibility is that the timing differences arise from the dynamics associated with an arcade of loops, rather than particle kinematics within a single flaring loop.  
In this scenario, flaring loops would sequentially brighten, with the white light flare emission occurring in newly formed flare kernels in the arcade, while the radio emission
is responding to loops filled earlier in the process.
This is observed in the Sun when the loop footpoints are seen to brighten sequentially along a polarity inversion line, with the distance between them growing with time, a signpost of magnetic reconnection proceeding to higher overlying loops \citep{Kowalski2024}. 
This scenario is compatible with co-occuring with the second scenario described above, but more detailed modelling would be needed to work out the expected timing for the two flare emissions. 

\subsection{The Properties of Optical Flares Occurring with and without a Radio Response \label{disc:flare_prop}}
This dataset reveals an apparent higher optical flare rate than radio flare rate.
Some of this undoubtedly reflects the underlying sensitivity of the radio data. 
While we do not have enough data to construct a flare-frequency distribution (FFD) for each wavelength region, we note that
\citet{Tristan2025} determined that radio, optical, UV, and X-ray FFDs for another active M dwarf, AU Mic, displayed similar overall distributions in terms of the index of the flare frequency distribution ($\beta$), even as the flare-to-flare behavior revealed dissimilarities in the wavelength-dependent flare responses. 
Thus clearly while the ensemble of flare processes appears to be the same, there can be significant flare to flare differences, as found in this study. 

The four optical flares with radio counterparts do not exhibit a consistent pattern of impulsiveness index
$\mathcal{I}$.
As discussed in \S~\ref{sec:spec+time} and \citet{Kowalski2016} the optical photometric variations alone cannot be used to predict the amount of the Balmer jump observed in the blue-optical portion of the flare spectrum. 
This also translates to an inability to determine based solely on photometry what the radio association for a given optical flare will be. 

Using the optical spectroscopy available on the night of 10 Oct. 2009 we examine the spectrum closest to the peak time of the optical photometry. Making use of recently available
radiative hydrodynamic models to explain the optical spectral energy distribution, we see that the two optical flares with accompanying radio responses are consistent with beam heating having high low energy cutoffs. 
The radio observations are generally sensitive to high energy electrons ($\sim$MeV-level).
This result needs to be confirmed with additional data examining the relationship between optical flare properties and
co-occurrence of radio flares, but suggests that flares with a response at radio wavelengths may have a higher average electron energy due to a higher low energy cutoff. 

The optical photometric properties of the flares here do not show a systematic behavior that would indicate ability to predict which optical flares would have a radio counterpart. 
While the sample size of optical flares with time-resolved spectroscopy and radio detections is small, the results here suggest that using the model inferences on the low energy cutoff of the electron distribution which can reproduce
the blue-optical spectra may be a more useful approach to understanding stellar flare particle acceleration at both radio and optical wavelengths.


\subsection{Radio Flares: Frequency-dependent delay in radio flare peaks \label{disc:radio_flares}}

There has to date been very little in the way of detailed observations or theoretical modelling of M dwarf radio flares. 
In this section we call attention to the curious frequency behavior of  two big radio flares discussed in \S~\ref{sec:multiw}. 
While both exhibited a delay in the maximum flare flux density with decreasing frequency,
the behavior in the decay phase was not the same. 
The 2001 Sept. 20 flare exhibited a frequency-dependent decay timescale which evolved over the course of the flare decay, while 
R1 has roughly the same decay timescale at the two frequencies after the time of peak 6 cm flux. 
We also note that such frequency-dependent delays in maximum flare flux density  with delays increasing to smaller frequencies have also been noted in some other M dwarf flares with multi-frequency coverage as well (L. Vega, priv. comm.). 
This is reminiscent of a notable solar flare diagnosed in \citet{Bastian2007} which
displayed a flux density maximum delayed for decreasing frequency and a decay timescale independent of radio frequency.
\citeauthor{Bastian2007} attribute this to the influence of free-free absorption from a dense plasma.
As we do not have multi-wavelength data to add context to radio flare R1, we can only speculate on possible plasma conditions and explore if the frequency-dependent delay could be due to extra opacity from free-free sources. 
Time-resolved spectroscopy would be ideal in this case for probing the velocity and characteristics of the 
chromospheric evaporation front (which should emit X-ray and UV emission from the heated gas, as surmised in \citet{Chen2021}).

\begin{figure}
    \plotone{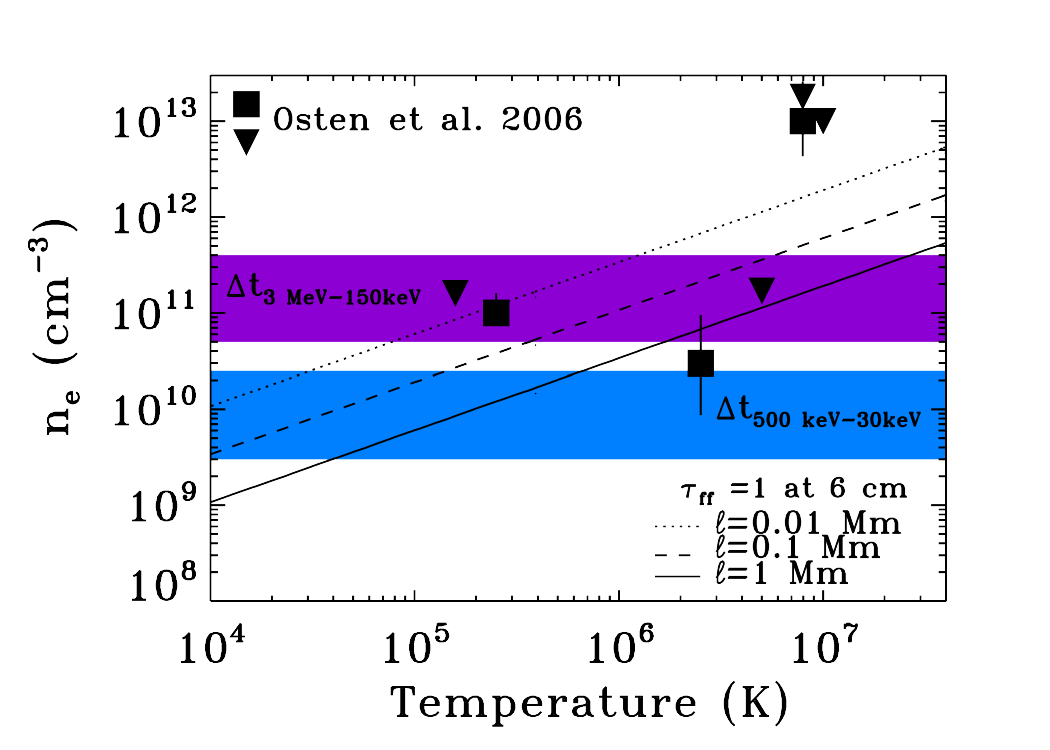}
    \caption{Contours of free-free optical depth as a function of electron density and temperature for three characteristic length scales of the path integral along a loop. 
    The area below and to the right of each line indicates the regions consistent with the material producing $\tau_{\rm ff}\leq 1$ at a wavelength (frequency) of 6 cm (4.8 GHz).
    Also overplotted in squares and downward triangles are electron density and temperature pairs (or electron density upper limits) determined from quiescent conditions of EV~Lac in \citet{Osten2006a}. The colored  regions indicate the range of electron densities where the observed time delay between optical and radio flare peaks is compatible with a difference in collisional deflection timescale between 500 keV-30 keV (blue) and 3 MeV-150 keV (purple); see Figure~\ref{fig:tau_defl} and discussion in \S~\ref{disc:opt+radio}.
    \label{fig:tauff}
    }
\end{figure}

A dimensional analysis of where in the ($n_{e}$,T) plane there would be sufficient free-free opacity at 4.8 GHz is revealing. The equation for free-free absorptivity at radio wavelengths is \\
\begin{equation}
\kappa_{\rm ff} \approx 0.2 n_{e}^{2} T_{e}^{-1.5}\nu^{-2}
\end{equation}
and indicates more absorptivity at smaller frequencies, and for the receiver bands used here (3.6 cm = 8.4 GHz, 6 cm = 4.8 GHz), there is roughly a factor of 3 higher absorptivity at 6 cm. 
For a homogeneous source the optical depth
$\tau = \kappa l$ for $l$ the appropriate length scale. 
There would be a narrow range of parameter space where such an explanation could hold, as moving to regions where $\tau$ is too high would lead to free-free absorption of most of the flaring radio emission (including at the higher frequency), and too low the effect would not be observable. 
Figure~\ref{fig:tauff} plots contours of $\tau_{\rm ff}$ versus $n_{e}$ and $T_{e}$. 
Also overplotted are measurements of
($n_{e}$,$T_{e}$) from \citet{Osten2006a}
taken from quiescent FUV and X-ray spectra of EV~Lac. 
These show that plausible plasma parameters in the low corona of EV~Lac exist to 
produce such a situation.
Apart from whether this flare produced a response at X-ray wavelengths (which we cannot constrain), the quiescent coronal conditions should persist. 
This could potentially be used as a diagnostic of
coronal conditions in future M dwarf flares.

The notable flare in \citet{ostenetal2005} also displayed this frequency-dependent behavior, and had an accompanying white light flare which peaked $\approx$54 seconds beforehand. 
The range of electron densities where this delay  can obtain from \S~\ref{disc:opt+radio} and Figure~\ref{fig:tau_defl} for two combinations of electron energies representing optical and radio-emitting electrons
is overplotted in Figure~\ref{fig:tauff}, indicating that for increased opacity from a chromospheric evaporation front to produce the observed behavior, implies path lengths of order 0.01-1 Mm at temperatures consistent with the transition region or below (for the less energetic combination of electron energies) 
or temperatures in the transition region to low corona. 
Note that this path length is not a loop length.
\citet{kowalski2015} compute the total physical depth 
of the dense chromospheric condensation where the 
precipitation producing the white-light flares occurs, at 20 km (=0.02 Mm). 
Momentum balance should be in effect between the downward-flowing chromospheric condensation (cc) and the upward-flowing chromospheric evaporation front (ef); with velocity ($v$), mass density ($\rho$), and depth range ($\Delta z$) for each component, \\
\begin{equation}
    (v \rho \Delta z)_{cc} \sim - (v \rho \Delta z)_{ef}
\end{equation}
and density contrast $\rho_{cc}/\rho_{ef}$ of an order of magnitude, and velocity ratio $v_{cc}/v_{ef}$ of about a factor of 3 \citep{Kowalski2024}, this implies a physical depth range for the evaporation front
of about 0.6 Mm, consistent with the path lengths constrained above.

We note that some preliminary modelling of stellar radio flares has been done, combining the radiative hydrodynamic modelling outputs described here in \S~\ref{sec:spec+time}
which can reproduce white light stellar flares with gyrosynchrotron emissivities and optical depth modelling using different magnetic geometries.
Tristan et al. (in prep.) were specifically motivated to reproduce the spectral index behavior of the 2001 Sept. 20 flare of EV~Lac, notably the flattening
 of the spectral index between 6 and 3.6 cm slightly before the time of peak 3.6 cm flux density. 
A large beam flux of 10$^{13}$ 
erg cm$^{-2}$ s$^{-1}$, electron distribution power-law index of $\delta=3$, and three different magnetic geometries (constant along the loop, exponentially decaying, or a bottled model with little change in B-field value until very close to the photosphere) were explored, with several configurations capable of reproducing the spectral index behavior, although the overall flux densities from the model were roughly three orders of magnitude too small to match the observed values.  
Future modelling attempts should take into account both gyrosynchrotron emissivities as well as potential effects of the
evaporating chromospheric plasma in modelling stellar radio flares.

\section{Conclusion \label{sec:conc}}
The present study has used simultaneous optical and radio data of M dwarf flares to study the correspondence of the action of accelerated electrons in two distinct regions of the stellar atmosphere.
This is analogous to the many similar studies performed on solar flares with hard X-ray and radio data, although utilizing observational capabilities and associated constraints which exist for the stellar case. 

We describe a sensitivity analysis to explore correlated multi-frequency behavior in the radio light curves, which can pick out smaller enhancements than relying on signal-to-noise ratio in one receiver band alone. 
From an analysis of these radio flares under the assumption that the electrons responsible for the emission are experiencing strong diffusion in a magnetic trap, we find evidence of extended structures, with heights 3-4 stellar radii above the photosphere, and 
low density environments, with $n_{e} <$ a few times $10^{8}$ cm$^{-3}$, and  $<10^{9}$ cm$^{-3}$
in two separate cases. 

The simultaneous data reveal a higher rate of optical flaring than radio flaring, although the limitations of the receiver bandwidth available for these radio data
are a contributing factor. Out of 9 optical flares seen over the course of 2 nights, only 4 had a detectable radio response. We find that optical photometric flare properties are not correlated with radio response.
From an analysis of the smaller number of optical flares with time-resolved low-spectral resolution data, we find suggestive evidence that the optical flares with a radio response occur as a result of beam heating produced from an electron distribution with a higher low energy cutoff than the optical flares without a radio response. 

Inspection of the detailed temporal behavior of the four events with both a radio and optical response reveal consistent timing patterns, in which the optical flare peaks before the microwave emission peaks. 
Taken together with an earlier flare recorded at radio and optical wavelengths, this time interval span $\sim$1-7 minutes, although data gaps due to calibration scans introduce uncertainties into the upper limit of this range.
Under the assumption that the radio emission originates from more energetic electrons in more rarefied environments than the white light emission, we find a few plausible explanations for the magnitude of this trend.  
One lies in different timescales for acceleration of low- vs high-energy accelerated electrons, rather than assuming all electrons are accelerated at the same time and place. 
Another explanation invokes the trapping timescales of the radio-emitting electrons as opposed to the directly-precipitating white light-emitting electrons, and we find a plausible range of electron densities to explain the observed timescales for representative electron energies undergoing collisional deflection. 
The third explanation relaxes the assumption about the flare dynamics taking place within a single magnetic loop, with sequential brightening in an arcade of loops.  
The reality is likely a complicated convolution of all three of these effects. 

We describe a radio flare in our dataset which exhibited a frequency-dependent delay in the peak flare flux, in which the 6 cm emission peaked about two minutes after the  3.6 cm emission peak. 
This behavior is similar to that noted in an earlier large radio flare.  We discuss the possibility that this frequency-dependent delay could be due to the temporary influence of free-free opacity from a chromospheric evaporation front in the early phases of the flare. 
From a dimensional analysis of densities and temperatures required to  achieve this effect,
we find a concordance with the densities implied to explain the observed time delay between optical and radio flares, as well as a consistency between the order of magnitude estimations of the size scale of the 
opaque region and that of a chromospheric evaporation front. 

One notable boon to stellar astrophysics in the last decade has been the increasing availability of sensitive optical photometric time-series, largely to increase discoveries of transiting extrasolar planets around nearby stars. 
This plus the recognition of the impact of stellar activity on 
the existence and make-up of exoplanet atmospheres has re-awakened
interest in multi-wavelength  time-domain observations of cool stars. 
The upgraded receivers of the JVLA enable up to 4 GHz of instantaneous bandwidth, compared with only 100 MHz available in this data.
So a new generation of optical-radio flare studies should reveal even more than what has been found in this study.
Some pertinent characteristics of the observing should be noted, in order to make advances in understanding the details of particle acceleration in a stellar environment. 
Given the timing differences noted in the current study, having similarly cadenced optical data as the radio data (of order 10s or less)
will be key to further probing of any 
systematic differences between trapped and precipitating electrons. 
This study has additionally shown the importance of low-resolution time-resolved spectroscopy capable of covering the blue-optical spectral region, 
in order to more precisely diagnose the accelerated electrons giving rise to the optical flare emission.
Multi-frequency radio observations are also key to diagnosing dynamical effects and potentially constraining the properties of a chromospheric evaporation front in the course of the flare. 
In addition to these observational desirements, advancements in models of stellar radio emission during flares are a requirement to 
move beyond dimensional analysis and place stellar radio astrophysics
on a more similar footing to what can be done with solar radio observations and interpretation.

\begin{acknowledgments}
We wish to thank John Wisniewski and Praveen Kundurthy for their help with acquiring the optical data. 
\end{acknowledgments}




\end{document}